\documentclass[aps,pra,twocolumn,showpacs,showkeys,superscriptaddress,reprint]{revtex4-1} 
\usepackage{graphics} 
\usepackage{graphicx} 
\usepackage{amsmath} 
\usepackage{color}

\begin{document} 

\title{Spin-selective insulators in Bose-Fermi mixtures} 
\author{R. Guerrero-Suarez}
\affiliation{Departamento de F\'{\i}sica, Universidad Nacional de Colombia, A. A. 5997 Bogot\'a, Colombia.}
\author{J. J. Mendoza-Arenas}
\affiliation{Departamento de F\'{\i}sica, Universidad de los Andes, A. A. 5997 Bogot\'a, Colombia.}
\author{R. Franco} 
\affiliation{Departamento de F\'{\i}sica, Universidad Nacional de Colombia, A. A. 5997 Bogot\'a, Colombia.}
\author{J. Silva-Valencia} 
\email{jsilvav@unal.edu.co} 
\affiliation{Departamento de F\'{\i}sica, Universidad Nacional de Colombia, A. A. 5997 Bogot\'a, Colombia.} 

\date{\today} 

\begin{abstract}
We investigate an imbalanced mixture composed of two-color fermions and scalar bosons in the hard-core limit, considering repulsive and attractive interspecies and intraspecies interactions. The interplay between commensurability, repulsive interactions and imbalance generates three insulating phases: a mixed Mott state and two spin-selective insulators characterized by the commensurability relations $\rho_B+\rho^{\uparrow,(\downarrow)}_F=1$. For an attractive coupling between fermions and bosons, we found the relations $\rho_B-\rho^{\uparrow,(\downarrow)}_F=0$ for the spin-selective insulators. State-of-the-art cold-atoms setups constitute ideal platforms to implement these unveiled insulating states and verify their commensurability relations.
\end{abstract} 


\maketitle 

\section{\label{sec1}Introduction}

The rapid development of the cold-atom field  has allowed researchers to verify and extend several predictions of condensed matter physics in clean and fully controllable setups~\cite{IBloch-RMP08,Esslinger-AR10,IBloch-NP12,Gross-S17}. 
Seminal examples include the observation and manipulation of superfluid-insulator phase transitions confining bosonic or fermionic isotopes of atoms~\cite{Greiner-N02a,Jordens-N08,Schneider-S08,Bakr-S10,Sherson-Nat10,Greif-Science16}. 
Following this path, a peculiar system exquisitely turned into reality in cold-atoms setups is that of mixed carriers which obey the Bose-
Einstein and Fermi-Dirac statistics~\cite{Truscott-S01,Schreck-PRL01,Hadzibabic-PRL02,Roati-PRL02,Ott-PRL04,Silber-PRL05,Gunter-PRL06,Ospelkaus-PRL06,Zaccanti-PRA06,McNamara-PRL06,Best-PRL09,Fukuhara-PRA09b,Deh-PRA10,Tey-PRA10,Sugawa-NP11,Schuster-PRA12,Tung-PRA13,Ferrier-Barbut-S14,Delehaye-PRL15,Vaidya-PRA15,XCYao-PRL16,Onofrio-PUsp16,YPWu-JPB17,Roy-PRL17,Schafer-PRA18}. Controlling the number of each kind of carriers, the inter and intraspecies interactions, it has been possible to evidence exciting phenomena such as phase separation~\cite{Lous-PRL18} and a Bose-Fermi superfluid mixture~\cite{Trautmann-PRL18}.\par 
It is common to describe a mixture of bosonic and fermionic atoms within the framework of the Bose-Fermi-Hubbard model, which has been addressed considering different approaches that have been improved over time.  Given a particular approximation of this model, several analytical and/or numerical techniques can be used, which predict the emergence of superfluids, charge density wave, Mott insulators, spin density wave, phase separation, Wigner crystals among other ground states~\cite{Albus-PRA03,Cazalilla-PRL03,Lewenstein-PRL04,Mathey-PRL04,Roth-PRA04,Frahm-PRA05,Batchelor-PRA05,Takeuchi-PRA05,Pollet-PRL06,Mathey-PRA07,Sengupta-PRA07,Mering-PRA08,Suzuki-PRA08,Luhmann-PRL08,Rizzi-PRA08,Orth-PRA09,XYin-PRA09,Sinha-PRB09,Orignac-PRA10,Polak-PRA10,Mering-PRA10,Anders-PRL12,Masaki-JPSJ13,Bukov-PRB14,TOzawa-PRA14,Bilitewski-PRB15}. Specifically, for the superfluid-insulator transitions we know that regardless of the sign of the boson-fermion interaction and for a fixed fermionic density $\rho_F$, there are always two non-trivial insulator phases between the trivial insulators at integer bosonic densities $\rho_B$. These non-trivial insulators satisfy the conditions $\rho_B\pm\rho_F=n$ and $\rho_B\pm \tfrac{1}{2}\rho_F=n$, ($n$ integer), with the  plus (minus) sign for repulsion (attraction)~\cite{Zujev-PRA08,Sugawa-NP11,Avella-PRA19,Avella-PRA20}. The latter condition characterizes the noncommensurate insulators whose origin will be discussed in the present article.\par 
The high degree of control over cold-atom setups has allowed experimentalists to generate asymmetries in the spin populations~\cite{Zwierlein-S06,Partridge-S06,Liao-Nat10,Kinnunen-RPP18,Dobrzyniecki-AQT20}, making such systems ideal for observing the elusive unconventional pairing mechanism named after Fulde, Ferrell, Larkin, and Ovchinnikov (FFLO)~\cite{Fulde64,Larkin65}. However, the effect of spin population imbalance on Bose-Fermi mixtures has been barely explored. In this direction, Singh and Orso very recently showed that the visibility of the FFLO state is enhanced as the boson-fermion repulsion grows~\cite{Singh-PRR20}. In the present article, we focus on the superfluid-insulator transitions and on the effect of the spin population imbalance on their location. Considering a mixture composed of two-color fermions and scalar bosons in the hard-core limit, we found that in presence of the imbalance, the noncommensurate insulator state is divided into two spin-selective insulator states, which are separated by a superfluid phase. These new incompressible states fulfill the relations $\rho_B \pm \rho^{\uparrow,(\downarrow)}_F=n$, where plus and $n=1$ (minus and $n=0$) correspond to a repulsive (attractive) boson-fermion interaction. In these spin-selective  insulators, one-color fermions satify a  commensurability relation with the bosons, while the others remain in a gapless phase, being this the main result of the present investigation. \par
The rest of this article is organized as follows. In Sec.~\ref{sec2} we introduce the Bose-Fermi Hubbard Hamiltonian considered in this investigation. Using the DMRG algorithm, we build several zero-temperature phase diagrams, which are shown in Secs.~\ref{sec3} and ~\ref{sec4} for repulsive and attractive fermionic interactions respectively. Our main findings are summarized in Sec.~\ref{sec5}.
\section{\label{sec2}Model}
A mixture of bosonic and fermionic atoms confined in a one-dimensional optical lattice can be described considering a Hubbard-like Hamiltonian for each species and an interspecies interaction characterized by the parameter $U_{BF}$, which can be 
attractive or repulsive. Therefore, the Bose-Fermi-Hubbard Hamiltonian is given by 
\begin{equation}\label{eq:HBF}
	\hat{H}_{BF}=\hat{H}_B+\hat{H}_F+U_{BF}\sum^{L}_{i=1}\hat{n}_i^B\left(\hat{n}_{i,\uparrow}^F+ \hat{n}_{i,\downarrow}^F\right)\text{.}
\end{equation}
\noindent Here $\hat{H}_B$ corresponds to
\begin{equation}\label{eq:HB}
	\hat{H}_B = -t_B \sum_{\langle i,j\rangle}\left(\hat{b}_i^\dagger\hat{b}_j + \text{h.c.} \right),
\end{equation}
\noindent which is the Bose-Hubbard Hamiltonian in the hard-core limit, i.e. the number of bosons is at most one per site and they do not interact with each other. Operator $\hat{b}_i^\dagger$ ($\hat{b}_i$) creates (annihilates) a boson at node $i$ of a lattice of size $L$, while $\hat{n}_i^B=\hat{b}_i^\dagger\hat{b}_i$ is the local number operator. The hopping of bosons between nearest-neighbor sites ($\langle i,j \rangle$) is modulated by the parameter $t_B$, $N_B$ is the number of bosonic atoms, and $\rho_B=N_B/L$ is the global density of bosons, which varies from zero to one.\par 
The Fermi-Hubbard Hamiltonian characterized by the hopping parameter $t_F$ and the local interaction $U_{FF}$ is
\begin{equation}\label{eq:HF}
\hat{H}_F = -t_F \sum_{\langle i,j\rangle,\sigma}\left(\hat{f}_{i\sigma}^\dagger\hat{f}_{j\sigma} + \text{h.c.} \right) + \frac{U_{FF}}{2}\sum_{i,\sigma\neq\sigma'}\hat{n}_{i,\sigma}^F\hat{n}_{i,\sigma'}^F.
\end{equation}

Here, $\hat{f}_{i,\sigma}^\dagger$ $(\hat{f}_{i,\sigma})$ creates (annihilates) a fermion with spin $\sigma=\uparrow,\downarrow$ at node $i$, and $\hat{n}^{F}_{i,\sigma}=\hat{f}_{i,\sigma}^{\dag}\hat{f}_{i,\sigma}$ is the local number operator for each kind of fermions. We define $\rho^{\sigma}_F=N^{\sigma}_F/L$ as the global density for $\sigma$-fermions, where $N^{\sigma}_F$ is the number of fermions with spin $\sigma$. Then the total fermionic density is $\rho_F=\rho^{\uparrow}_F+\rho^{\downarrow}_F$, which  varies from zero to two, $\rho_F=1$ being the half-filling configuration. The spin population imbalance is quantified by the parameter $I=(N_F^\downarrow-N_F^\uparrow)/
(N_F^\uparrow+N_F^\downarrow)$. By fixing $t_F= t_B= 1$, we establish our energy scale.\par
Diverse mixtures of bosonic and fermionic isotopes have been achieved in cold atom setups using the same and different atoms, although their stability is severely limited by three-body recombinations~\cite{Sowinski-RPP19}. The model considered in this paper can be emulated in current cold atom setups; for instance the isotopes $^{171}$Yb and $^{174}$Yb ($^{170}$Yb) can be used~\cite{YTakasu-JPSJ09}. Also, we highlight that dual Bose-Einstein condensates of paired fermions and bosons with $^6$Li and $^7$Li have been achieved experimentally~\cite{Ikemachi-JPB17}.
\par
\begin{figure}[t] 
\includegraphics[width=18pc]{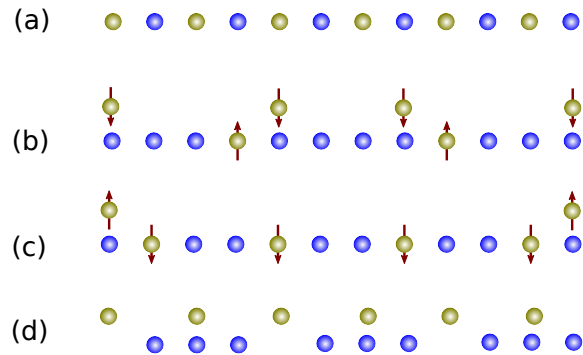} 
\caption{\label{fig1} Sketches of possible distributions of carriers of a mixture of two-color fermions and scalar bosons in a chain of lenght $L=12$. Blue (golden) circles represents bosons (fermions). A mixed Mott insulator state with $\rho_F = 1/2$ and $\rho_B = 1/2$, for which $\rho_F+\rho_B=1$, is depicted in (a). Possible ground states for an imbalance mixture with $I=1/3$ are shown in (b) and (c) such that the relations $\rho^{\uparrow(\downarrow)}_{F}+\rho_B=1$ are fulfilled, respectively. Without imbalance, a noncommensurate state emerges (d), which satisfies $\rho_B+\tfrac{1}{2}\rho_F=1$. Note that from the above state, the configurations (b) and (c) arise in the presence of imbalance and when the number of bosons increases and decreases by one, respetively.}
\end{figure} 
It is well known that the commensurability condition, i.e. the fact that the number of carriers is proportional to the lattice size, is critical for the emergence of insulating states in pure bosonic and fermionic systems~\cite{Cazalilla-RMP11,Guan-RMP13}. For instance, two-color fermions in one dimension exhibit a Mott insulator phase for $\rho_F=1$, and insulator phases emerge at integer densities in bosonic systems. When bosons and fermions coexist, this commensurability condition also arises. This corresponds to the case where the total number of carriers (bosons + fermions) matches the lattice size, establishing the relation $\rho_B+\rho_F=1$ between the global bosonic and fermionic densities. This scenario was observed in experiments with Yb atoms~\cite{Sugawa-NP11}, and predicted for polarized carriers~\cite{Zujev-PRA08} and a mixture of scalar bosons with two-color fermions~\cite{Avella-PRA19,Avella-PRA20}. In  Fig. ~\ref{fig1} (a), we sketch the mixed Mott insulator state for which $\rho_B+\rho_F=1$, while commensurate insulators consisting of bosons and one kind of fermions are shown in Fig. ~\ref{fig1} (b) and (c). An ilustration of a noncommensurate insulator is sketched in  Fig.~\ref{fig1} (d), which satisfies the relation $\rho_B+\tfrac{1}{2}\rho_F=1$.\par
To explore the effect of the spin population imbalance on the superfluid-insulator transitions of a mixture of two-color fermions and scalar bosons, we consider separately the attractive and repulsive interactions between fermions, and particular densities of the latter. 
The ground-state energy for $N_B$ bosons, and $N^{\uparrow}_F$ and $N^{\downarrow}_F$ fermions, denoted by  $E(N^{\uparrow}_F,N^{\downarrow}_F,N_{B})$, was calculated using the density-matrix renormalization group with open boundary conditions~\cite{White-PRL92}. Namely, we used a local basis with eight states: $\frac{|F\rangle}{|B\rangle}=\frac{|0\rangle}{|0\rangle},\frac{|\uparrow\rangle}{|0\rangle},\frac{|\downarrow\rangle}{|0\rangle},\frac{|\uparrow\downarrow\rangle}{|0\rangle},\frac{|0\rangle}{|1\rangle},\frac{|\uparrow\rangle}{|1\rangle},\frac{|\downarrow\rangle}{|1\rangle},\frac{|\uparrow\downarrow\rangle}{|1\rangle}$, and considered the dynamical block selection state protocol~\cite{Legeza-PRB03}, which allows us to set up the DMRG truncation error, while the number of maintained states varies. We kept a discarded weight of $\approx10^{-7}$ and fixed the minimum number of states to 400, while the algorithm considered a maximum of 1800 states. To obtain a energy convergence of $10^{-3}$ or lower, we carried out between nine or eleven sweeps.\par 
\begin{figure}[t]
	\centering
	\includegraphics[width=19pc]{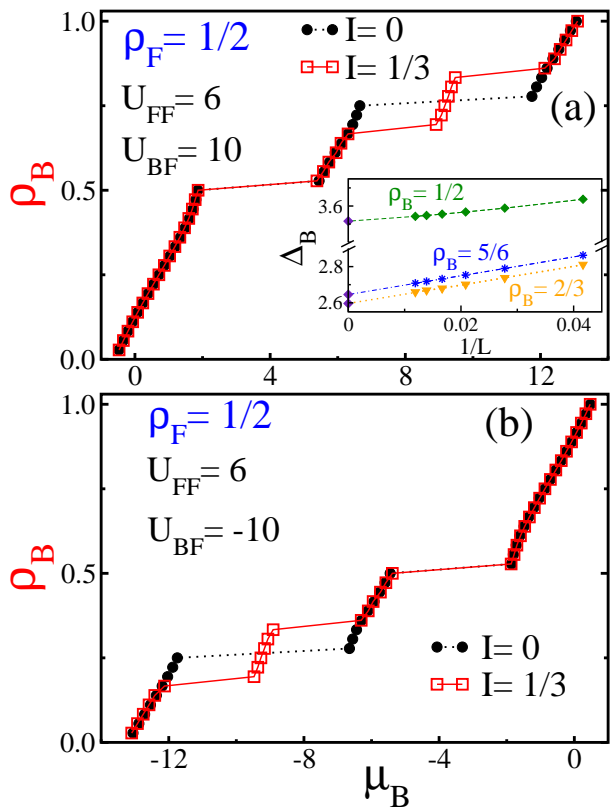}
	\caption{Bosonic density $\rho_B$ as a function of bosonic chemical potential $\mu_B$, calculated in the thermodynamic limit, for fermionic density $\rho_F=1/2$ and repulsive fermion-fermion interaction $U_{FF}=6$.	Two different imbalance values were considered, namely $I=0$ (black dots) and $I=1/3$ (red squares). (a) Repulsive inter-species interaction $U_{BF}=10$. Inset: Charge gap $\Delta_B$ as a function of the system size for the bosonic densities leading to insulating states. The diamonds on the $y$ axis correspond to the extrapolations to the thermodynamic limit. (b) Attractive inter-species interaction $U_{BF}=-10$. The lines are visual guides}
	\label{fig2}
\end{figure}
\section{\label{sec3}Repulsive Fermionic Interactions $U_{FF}>0$}
We start by studying the scenario where fermions repel each other, and we focus on the densities of quarter- and half- fermionic filling. However the main results were corroborated for other densities, imbalance and interaction parameters.\par
\subsection{Quarter-Filling Case $\rho_F=1/2$}
In this section, we consider a mixture of two-color fermions and scalar bosons with a global density of fermions of $\rho_F=1/2$, and a repulsive fermionic interaction $U_{FF}=6$, parameters that will remain fixed. For both positive and negative values of the boson-fermion coupling ($|U_{BF}|= 10$), we display in  Fig.~\ref{fig2} the evolution of the bosonic chemical potential 
$\mu_B=E(N^{\uparrow}_F,N^{\downarrow}_F,N_{B}+1)-E(N^{\uparrow}_F,N^{\downarrow}_F,N_{B})$ as the number of bosons increases from zero. For a spin-balanced mixture, i.e. equal number of fermions for each color ($I=0$), we see that the chemical potential mostly grows monotonously as the number of bosons increases, indicating that there is no cost to generate excitations for the majority of bosonic densities. However, this behavior changes for two particular densities. In the case of boson-fermion repulsion, these correspond to $\rho_B=1/2$ and $\rho_B=3/4$ (Fig. ~\ref{fig2} (a)). The former density is related to the mixed Mott insulator state for which $\rho_B+\rho_F=1$, and the latter is a noncommensurate insulator for which $\rho_B+\tfrac{1}{2}\rho_F=1$. The width of these plateaus when $L\to\infty$ corresponds to the charge gap $\Delta_B$ for bosonic excitations in the thermodynamic limit.
For different spin populations $N^{\uparrow}_F\neq N^{\downarrow}_F$, we observe that gapless and gapped states emerge as in the balanced case. Namely, we obtain that the mixed Mott insulator state always appears, remaining unchanged regardless of value of the imbalance parameter. This result is expected since the total number of carriers must be commensurate with the lattice size regardless of the asymmetry in the spin populations. In Fig.~\ref{fig2} (a) we consider $I=1/3$, and the main change in the $\rho_B$-$\mu_B$ curve is that the noncommensurate insulator splits into two insulators, which are separated by a superfluid region; this suggests a gapless region that can be a polaronic Luttinger liquid~\cite{Mathey-PRL04} (see the appendix). Then, for an imbalance of one-third, the plateaus are located at the bosonic densities $\rho_B=1/2$, $\rho_B=2/3$, and $\rho_B=5/6$. Note that the evolution of the width of these plateaus with the lattice size leads to finite gaps in the thermodynamic limit, namely $\Delta_B^{\rho_B=1/2}=3.52$, $\Delta_B^{\rho_B=2/3}=2.61$ and $\Delta_B^{\rho_B=5/6}=2.65$, which were obtained using a second-order polynomial extrapolation (see inset of Fig. ~\ref{fig2} (a)). As the imbalance increases the new plateaus move away from the noncommensurate one, and for the extreme imbalance ($I_{max}=1$) they tend to the bosonic densities $\rho_B=1/2$ and $\rho_B=1$, recovering the spin polarized results, i.e. the trivial plateau and the mixed Mott insulator state~\cite{Zujev-PRA08}. The above information allows us to establish a relation between the global densities of the carriers and the imbalance to determine the location of the new insulator, which is $\rho_B+\frac{1}{2}\rho_F(1 \pm I)=1$. Notice that with $\rho_F=1/2$ and $I=1/3$, the insulators correspond to $\rho_B=2/3$ and $\rho_B=5/6$, thus obtaining the densities discussed above.\par 
\begin{figure}[t]
	\centering
	\includegraphics[width=21pc]{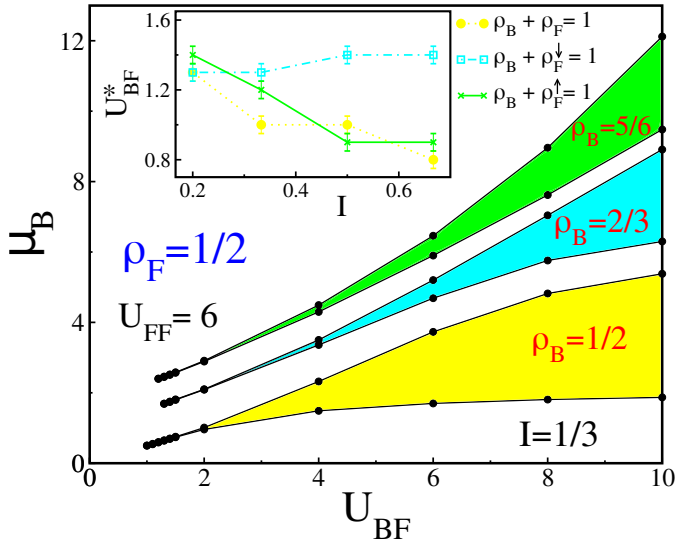}
	\caption{Phase diagram in the chemical potential vs inter-species interaction plain. The fermion-fermion interaction is $U_{FF}=6$, the fermion density is $\rho_F=1/2$ and the imbalance is $I=1/3$. The inset shows the dependence of the critical points $U_{BF}^*$ with the imbalance. The lines are visual guides; matching colors to those of the lobes in the main panel indicate that they correspond to the same bosonic density.
	}
	\label{fig3}
\end{figure}
For attractive inter-species interaction $U_{BF}=-10$ and without spin imbalance, the charge gap vanishes except at the bosonic densities $\rho_B=1/4$ and $\rho_B=1/2$. For these densities clear discontinuities appear, as shown in Fig.~\ref{fig2} (b), indicating that insulator states emerge. It is evident that the sign of the Bose-Fermi coupling modifies the insulator states of a mixture of scalar bosons and spinor fermions, as was reported in a recent paper by three of us in the soft-core approximation without population imbalance~\cite{Avella-PRA20}. There it was found that the positions of these noncommensurate insulator states satisfy the relations $\rho_B-\rho_F=0$ and $\rho_B-\frac{1}{2}\rho_F=0$. The former noncommensurate insulator is characterized by a local pairing of one fermion and one boson, leading to a global charge density wave state, where bosons and fermions are in phase. Furthermore, this state was reported experimentally by Sugawa \textit{et al.}~\cite{Sugawa-NP11}. Now, for any asymmetry between the spin populations, the plateau that fulfills the relation $\rho_B-\frac{1}{2}\rho_F=0$ splits into two new insulators ($\rho_B=1/3$ and $\rho_B=1/6$ for $I=1/3$, as shown in Fig.~\ref{fig2} (b)), while the rest of the curve remains unaltered. We establish that the location of the new insulators is given by the relation $\rho_B+\frac{1}{2}\rho_F(-1 \pm I)=0$, which for extreme imbalance leads to a trivial plateau at $\rho_B=0$ and other at $\rho_B=1/2$. Comparing the plots of 
Fig.~\ref{fig2}, we note that the plateaus are related by the transformation $\rho_B\rightarrow 1-\rho_B$ when going from a repulsive to an attractive Bose-Fermi coupling. We also notice that $| \mu^{A}_{B}\left(\rho_B\right)|=| \mu^{R}_{B}\left(1-\rho_B\right)|$, where $A$ and $R$ correspond to attractive and repulsive interactions respectively, and the charge gaps in the thermodynamic limit are the same regardless of the sign of $U_{BF}$.\par 
\begin{figure}[t]
	\centering
	\includegraphics[width=19pc]{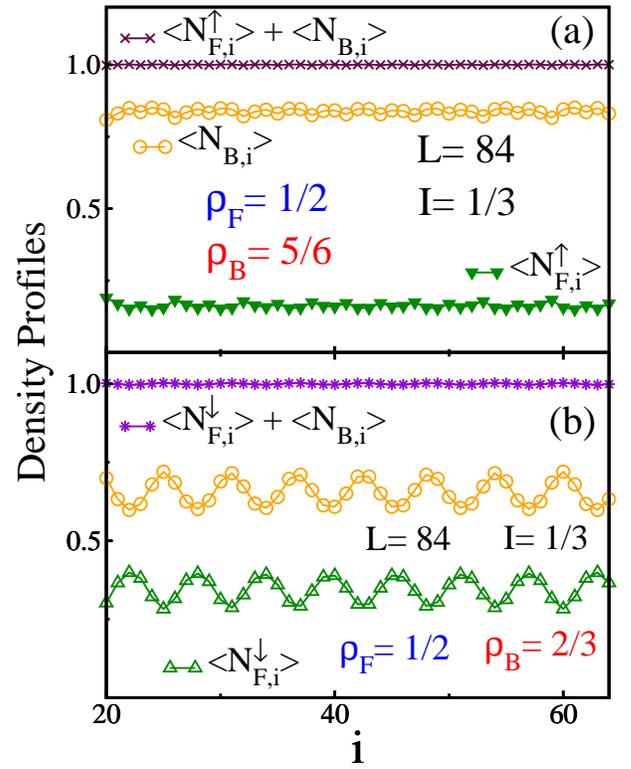}
	\caption{Density profiles of bosons and fermions for a system of size L = 84, with fermion-fermion repulsion $U_{FF}=6$, fermionic density $\rho_F=1/2$, inter-species interaction $U_{BF}=10$ and imbalance $I=1/3$. (a) Insulator state at $\rho_B=5/6$. (b) Insulator state at $\rho_B=2/3$. The lines are visual guides.}
	\label{fig4}
\end{figure}
\begin{figure}[t]
	\centering
	\includegraphics[width=19pc]{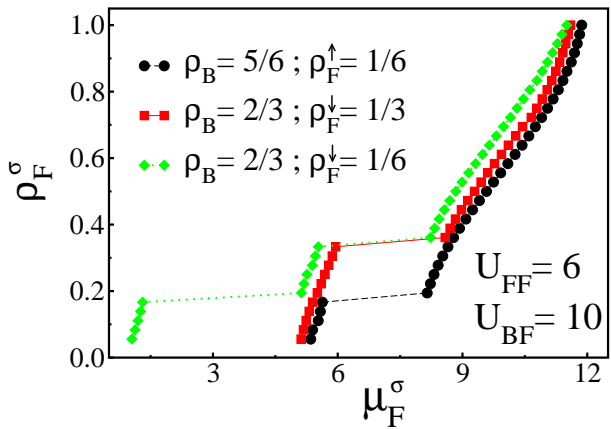}
	\caption{Fermionic density versus chemical potential for the variable kind of fermions $\sigma=\downarrow,\uparrow$. Black dots: The number of spin-down fermions changes, whereas the bosonic density $\rho_B=5/6$ and the density of spin-up fermions $\rho^{\uparrow}_{F}=1/6$ are fixed. In the other curves, the bosonic density is fixed to $\rho_B=2/3$ and the number of spin-up fermions varies, considering $\rho^{\downarrow}_{F}=1/3$ (red squares) and $\rho^{\downarrow}_{F}=1/6$ (green diamonds). Here, the interaction parameters are $U_{FF}=6$ and $U_{BF}=10$. The lines are visual guides.}
	\label{fig5}
\end{figure}
To explore the occurrence of insulating states for other values of the boson-fermion interation, we replicate Fig.~\ref{fig2} (a) for different $U_{BF}$ keeping the fermionic density ($\rho_F=1/2$), the fermion-fermion interaction ($U_{FF}=6$) and the spin population imbalance ($I=1/3$) constant. In this form we obtained the phase diagram shown in Fig.~\ref{fig3}, where the white regions correspond to superfluid states, i.e. the charge gap vanishes and the ground state is compressible. On the other hand, incompressible states, signaled by a finite gap in the thermodynamic limit, are represented by colorful regions. Note that the extreme value of this phase diagram ($U_{BF}=10$) corresponds to Fig.~\ref{fig2} (a). Three lobes separated by superfluid regions are clearly seen in this phase diagram. The yellow lobe corresponds to the mixed Mott state ($\rho_B=1/2$), while the others emerge from the imbalance, namely the cyan lobe for $\rho_B=2/3$ and the green lobe for $\rho_B=5/6$. Note that for a spin-balanced mixture the latter lobes combine to form a single lobe ($\rho_B=3/4$), corresponding to a noncommensurate state which dominates the phase diagram. As the Bose-Fermi coupling takes lower values, the charge gap decreases monotonously for all insulating lobes. Therefore they shrink and finally disappear at the critical values $U^{*}_{BF}=1.0$, $1.3$, and $1.2$ for $\rho_B=1/2, 2/3$, and  $\rho_B=5/6$, respectively; see the inset of Fig.~\ref{fig3}. Hence, below the critical points the ground state is superfluid. The fact that criticality occurs at distinct interactions for $\rho_B=2/3$ and $\rho_B=5/6$ reinforces that they constitute well-separated insulators . This is further emphasized by the different evolution of $U^{*}_{BF}$ as a function of the imbalance parameter, shown in the inset of Fig.~\ref{fig3}. For $\rho_B=2/3$ the critical point changes very little with $I$, while for the insulating lobe with $\rho_B=5/6$ the position of $U^{*}_{BF}$ goes to lower values. In addition, we observe that the critical point for the mixed Mott state decreases as the imbalance parameter grows, even though its density remains unaltered by the asymmetry in the spin-population. This can be understood by considering that a larger imbalance leads to a stronger effective repulsion of neighboring fermions, as their hopping to sites occupied by others of the same spin orientation is impeded, thus increasing the insulator lobe. Similar analyses were performed for the attractive Bose-Fermi coupling case, for which we obtained a phase diagram with three insulating lobes with very similar properties to those discussed above (not shown).\par 
\begin{figure}[t]
	\centering
	\includegraphics[width=19pc]{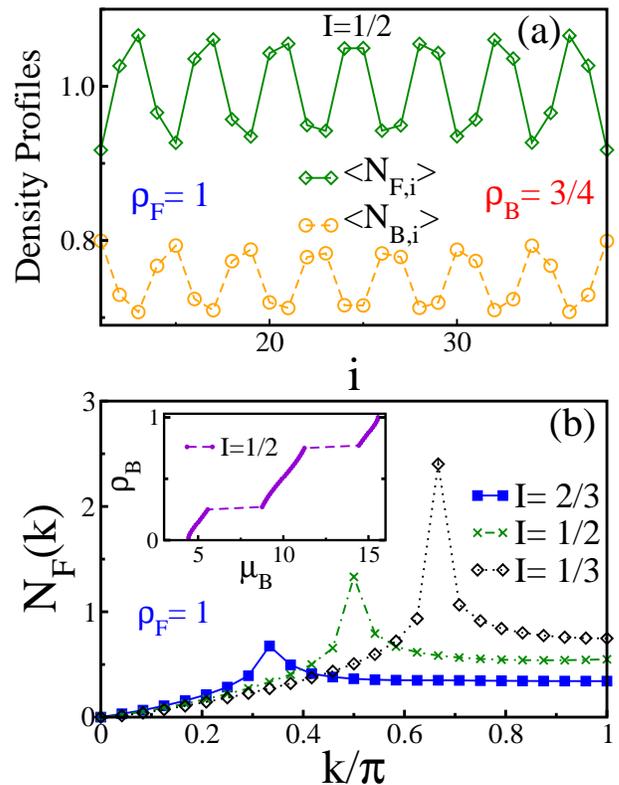}
	\caption{(a) Density profiles for an imbalanced mixture of bosons and fermions with a fixed fermionic density $\rho_F=1$, $I=1/2$, boson-fermion repulsion $U_{BF}=10$, fermion-fermion coupling $U_{FF}=6$, and a lattice size of $L=48$. A dimerized ground state is obtained for $\rho_B= 3/4$. (b) Fermion charge structure factor $N_F(K)$ as a function of the wave vector. Here, we consider $I= 1/3, 1/2$, and $2/3$. In the inset, we show the bosonic  density versus chemical potential for a mixture with $I=1/2$, clearly evidencing the emergence of two non-trivial plateaus. In both plots, the lines are visual guides.}
	\label{fig6}
\end{figure}
After establishing that the spin population imbalance splits the noncommensurate plateau into two new insulator phases, we discuss the associated distribution of carriers. In Fig.~\ref{fig4}, we display the density profiles of bosons and fermions along the lattice, considering the same global fermionic density and interaction parameters as in  Fig. ~\ref{fig2} (a). For the insulating state characterized by a bosonic density $\rho_B=5/6$, the expectation value of the local number of bosons slightly oscillates around $\left<N_{B,i} \right>\approx 0.833$, and a similar behavior along the lattice is observed for the expectation value of the local number of fermions with spin up $\left<N^{\uparrow}_{F,i} \right>\approx 0.166$ (see Fig.~\ref{fig4} (a)). Surprisingly, we note that locally the system satisfies $\left<N^{\uparrow}_{F,i} \right>+\left<N_{B,i} \right>=1$. While bosons and fermions with spin up adjust to meet local commensurability, fermions with spin down exhibit a local density expectation value $\left<N^{\downarrow}_{F,i} \right>\approx 0.333$. To clarify the role of the spin-down fermions in the insulator described above, we fixed the number of bosons and spin-up fermions, while the number of spin-down fermions varies. Our results are presented in Fig.~\ref{fig5}, where the evolution of the fermionic chemical potential $\mu^{\sigma}_F=E(N^{\sigma}_F+1,N^{\sigma'}_F,N_{B})-E(N^{\sigma}_F,N^{\sigma'}_F,N_{B})$ as a function of the density for one specific color is shown. Fixing $\rho_B=5/6$ and $\rho^{\uparrow}_F=1/6$, we obtained that the curve is continuous except for $\rho^{\downarrow}_F=1/6$ (Fig.~\ref{fig5}, black dots), consistent with the expectations for the balanced case $I=0$. Hence for a global fermionic density $\rho_F=1/2$ and an imbalance of $I=1/3$ ($\rho^{\downarrow}_F=1/3$), the spin-down fermions are in a gapless state. This already suggests that the incompressible state for $\rho_B=5/6$ is characterized by a combination of a gapless and an insulator state for fermions with spin down and up, respectively. Also, we unveil a new link between an insulating state and global  commensurability, i.e. this spin-selective state fulfills the relation $\rho_B+\rho^{\uparrow}_F=1$, which means that the number of bosons plus the number of fermions with spin up is commensurate with the lattice.\par
For the other incompresible state that emerges due to the imbalance, we obtained that all the density profiles exhibit a charge density wave structure along the lattice. Specifically we show the  expectation value of the local number of fermions with spin down $\left<N^{\downarrow}_{F,i} \right>$ and $\left<N_{B,i} \right>$ in Fig.~\ref{fig4} (b) for $\rho_B=2/3$. In this case, $\left<N^{\uparrow}_{F,i} \right>$ oscillates along the lattice (not shown), while the addition of $\left<N^{\downarrow}_{F,i} \right>$ and $\left<N_{B,i} \right>$ at each site is approximately 1, with both featuring oscillations of characteristic wave-vector $2k_{F_\downarrow}$. Hence the insulator state for $\rho_B=2/3$ incorporates the fact that the number of bosons plus the number of fermions with spin down is globally commensurated with the lattice ($\rho_B+\rho^{\downarrow}_F=1$), whereas the fermions with spin up remain in a gapless state because for $\rho^{\uparrow}_F=1/6$, the chemical potential in the thermodynamic limit is continuous (Fig.~\ref{fig5}, red squares). To evidence even further the emergence of the spin-selective insulators, we show in Fig.~\ref{fig5} the $\rho^{\uparrow}_F-\mu_F$ curve  for conditions distant from commensurability, i.e., here $\rho_B=2/3$ and $\rho^{\downarrow}_F=1/6$ (green diamonds). Increasing the number of spin-up fermions, the fermionic chemical potential is continuous until a discontinuity arises at  $\rho^{\uparrow}_F=1/6$, which determinates the mixed Mott insulator. Afterwards the chemical potential grows continuously and the spin-selective insulator emerges at $\rho^{\uparrow}_F=1/3$, i.e., when the number of spin-up fermions plus the number of bosons coincides with the lattice size, whereas the spin-down fermions remain in a gapless state. After verifying the same scenario for the incompressible state of $\rho_B=5/6$, we conclude that the imbalance and the repulsion between the fermions generate two spin-selective insulator states, which satisfy the commensurability relations $\rho_B+\rho^{\uparrow,(\downarrow)}_F=1$, in a gapless fermion polarized background. Note that an analogous physical idea explains the ferromagnetic metallic phase in the Kondo lattice model, where the majority-spin conduction electrons are metallic while the minority-spin electrons show insulating behavior, with the latter satisfying a commensurability relation with the localized spins~\cite{Peters-PRL12}.\par
Based on the above analysis, we revisit the balanced mixture case, where a noncommensurate insulator arises which satisfies the relation $\rho_B+\tfrac{1}{2}\rho_F=1$. However, fundamental information can be unveiled by considering that a balanced mixture corresponds to $\rho^{\uparrow}_F=\rho^{\downarrow}_F=\tfrac{1}{2}\rho_F$. Therefore bosons and any kind of fermions fulfill a commensurability relation, which is clear from Fig.~\ref{fig5}, where most of the plateaus correspond to the balance case ($I=0$). Hence to add one fermion to the mixture, a finite amount of energy must be paid. The imbalance thus leads to an asymmetry that splits the noncommensurate insulator.\par 
An identical analysis can be performed for an attractive boson-fermion interaction. As mentioned before, in this case and with no imbalance we found a noncommensurate insulator state characterized by the relation $\rho_B-\frac{1}{2}\rho_F=0$ \cite{Avella-PRA20}, which implies that one boson can couple locally with any kind of fermions generating an insulator. In an imbalanced Bose-Fermi mixture, two spin-selective insulator states will arise as before, which satisfy the relations $\rho_B-\rho^{\uparrow,(\downarrow)}_F=0$ (not shown). Note that in this form we recover our constraint condition $\rho^{\uparrow}_F+\rho^{\downarrow}_F=\rho_F.$ 
\begin{figure}[t]
	\centering
	\includegraphics[width=19pc]{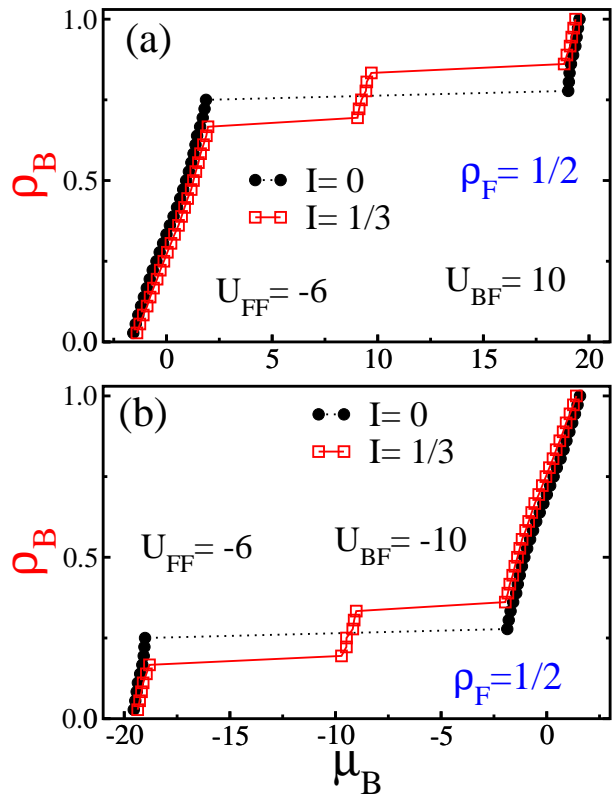}
	\caption{Bosonic density vs the thermodynamic-limit chemical potential for a Bose-Fermi mixture with attractive interaction between fermions ($U_{FF}=-6$). Here a balanced ($I=0$) and an imbalanced ($I= 1/3$) mixtures were considered with a fermion density $\rho_F=1/2$. Results for repulsive (a) and attractive (b) boson-fermion coupling are shown. The lines are visual guides.}
	\label{fig7}
\end{figure}
\subsection{Half-Filling Case $\rho_F=1$}
Commensurability conditions are fundamental for strongly correlated systems of bosons and fermions in one-dimension. For two-color fermions, this key condition takes place at $\rho_F=1$, and together with a repulsive interaction between fermions, they generate the Mott insulator state. Adding scalar bosons and turning on the  interspecies repulsion, it is expected that the mixed Mott state does not appear; instead, a noncommensurate insulator state emerges, which satisfies the relation $\rho_B+\tfrac{1}{2}\rho_F=1$~\cite{Avella-PRA19}. As discussed before, an asymmetry between the spin populations splits the above noncommensurate insulator into two spin-selective incompressible states, which are shown in the inset of Fig.~\ref{fig6} (b) for $I=1/2$. In this case, these correspond to $\rho_B=1/4$ and $\rho_B=3/4$, which are in agreement with the unveiled relation $\rho_B+\frac{1}{2}\rho_F(1\pm I)=1$. We remark that as before, the particular $\sigma$-mixed commensurability conditions $\rho_B+\rho_F^{\sigma}=1$ are fulfilled ($\sigma=\uparrow,\downarrow$), on top of a gapless polarized background.\par 
The density profiles for bosons and fermions along a chain of $L=48$ sites are presented in Fig.~\ref{fig6} (a), where  $I=1/2$, $U_{FF}=6$, and $U_{BF}=10$. In particular, we show the density profiles for the insulator state with $\rho_B=3/4$, although the same behavior is obtained for the other incompressible phase. The ground state exhibits a dimer structure for both bosons and fermions, which are out of phase. This is a particular distribution of carriers which already arose in previous studies on Bose-Fermi mixtures that considered fermionic half-filling~\cite{Titvinidze-PRL08}. The fact that both bosons and fermions exhibit a dimer distribution along the lattice indicates that the density-density correlations for both kind of carriers will have the same behavior. With this in mind, we display the charge structure factor for fermions 
\begin{equation}\label{eq:FSF}
\mathcal{N}^{F}(k)=\sum_{j,l=1}^{L}e^{ik(j-l)}\left(\langle n_{j}^{F} n_{l}^{F}\rangle-\langle n_{j}^{F}\rangle\langle n_{l}^{F}\rangle\right)
\end{equation}
\noindent in Fig.~\ref{fig6} (b) for $\rho_B=3/4$. Clearly a ground state with a dimer structure has a unit cell composed by four sites, which leads to a maximum of $\mathcal{N}^{F}(k)$ located at  $k=\frac{2\pi}{4}=0.5\pi$. Also, interwoven distributions of carriers are obtained for other values of the imbalance parameter, which are characterized by a maximum of the charge structure factor in a specific value of $k$. In Fig.~\ref{fig6} (b), we see that $\mathcal{N}^{F}(k)$ is maximum at $k/\pi=1/3$ and $2/3$ for imbalance $I=2/3$ and $1/3$, respectively. We conclude that special distributions of carriers take place for fermionic half-filling, which are marked by a maximum in the charge structure factor located at $k/\pi= 1-I$; these results are maintained for attractive interactions. For other fermionic fillings the distribution of carriers is more intricate, commonly with more than one broad peak in the charge structure factor, which cannot be described by such a simple relation.\par 
\begin{figure}[t]
	\centering
	\includegraphics[width=21pc]{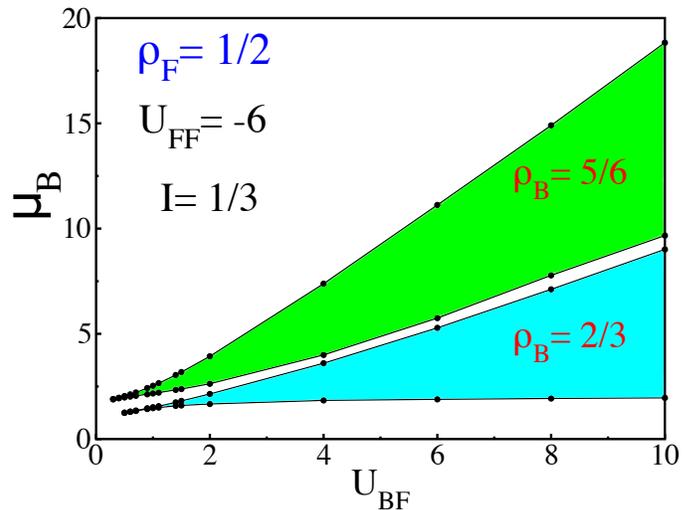}
	\caption{Bosonic chemical potential as a function of the inter-species interaction for an attractive coupling between fermions $U_{FF}= -6$. The fermion density is $\rho_F=1/2$ and the imbalance is $I=1/3$. The colourful regions indicate insulating phases with bosonic densities $\rho_B=2/3$ (yellow) and $\rho_B=5/6$ (cyan), whereas the white correspond to superfluid phases. The lines are visual guides.}
	\label{fig8}
\end{figure}
\section{\label{sec4}Attractive Fermionic Interactions $U_{FF}<0$}
In this section, we explore the superfluid-insulator transitions for a mixture of two-color fermions and scalar bosons considering an attractive interaction between the fermions. A fermionic density of $\rho_F=1/2$ was chosen and the absolute value of the interaction parameters match with those of Fig.~\ref{fig2} to directly compare the results. In Fig.~\ref{fig7} (a) we show the evolution of the bosonic chemical potential as the number of bosons increases from zero, for a repulsive interaction between bosons and fermions ($U_{BF}=10$). As the number of bosons grows in a balanced mixture, the charge gap in the thermodynamic limit vanishes for all densities except at $\rho_B=3/4$, where a strong discontinuity take place. The charge gap for this insulator state is $\Delta_B^{\rho_B=3/4}=17.15$. Surprisingly, the noncommensurate plateau survives and is related to the same condition found before, namely $\rho_B+\tfrac{1}{2}\rho_F=1$. On the other hand, the mixed Mott state, crucial for repulsive interactions, disappears when the coupling between fermions is attractive. This is sensible since the Mott state is induced by the commensurate condition and repulsive interactions. Comparing Fig.~\ref{fig2} (a) and Fig.~\ref{fig7} (a), we conclude that an attractive interaction between fermions prevents the mixed Mott state, but enhances the charge gap related to the noncommensurate insulator. The plateau corresponding to the latter state splits into two when an asymmetry between the spin populations is considered, in a similar way to the repulsive case. We establish that these new spin-selective insulator states satisfy the commensurability relations $\rho_B+\rho^{\uparrow,(\downarrow)}_F=1$, on top of  a gapless polarized background. This relation remains valid since the Pauli exclusion principle leads us to an effective repulsion between fermions of the same color, resulting in a scenario similar to that of Sec.~\ref{sec3}.\par 
In Fig.~\ref{fig7} (b), we show the $\rho_B$-$\mu_B$ curve for attractive interaction parameters, i.e. $U_{FF}=-6$ and $U_{BF}=-10$. Again, from the two noncommensurate insulating states one survives and its charge gap is enhanced ($\rho_B-\frac{1}{2}\rho_F=0$), whereas the other disappears ($\rho_B-\rho_F=0$). Here the effective repulsion due to the Pauli exclusion principle allows the emergence of the new spin-selective states ($\rho_B-\rho^{\uparrow,(\downarrow)}_F=0$) in presence of imbalance; this is exemplified for $I=1/3$. Comparing Figs. ~\ref{fig7} (a) and (b), we again establish that $| \mu^{A}_{B}\left(\rho_B\right)|=| \mu^{R}_{B}\left(1-\rho_B\right)|$ when going from a repulsive to an attractive boson-fermion interaction.\par
The zero temperature phase diagram for an attractive interaction between fermions of $U_{FF}=-6$ is shown in Fig.~\ref{fig8}, where an imbalance of $I=1/3$ and a fermionic density of $\rho_F=1/2$ have been considered. According to Fig.~\ref{fig7} (a), an insulating phase emerges at the bosonic density $\rho_B=3/4$, which splits into two spin-selective insulators in the presence of the imbalance. This scenario is maintained for other values of the inter-species interactions as can be seen in Fig.~\ref{fig8}, where the charge gap in the thermodynamic limit decreases with $U_{BF}$, determining insulating lobes for the bosonic densities $\rho_B= 2/3$ (cyan) and $\rho_B= 5/6$ (green), which are surrounded by superfluid regions (white zones). The critical inter-species interaction from which the spin-selective insulator arises is different for each lobe, namely $U^{*}_{BF}= 0.6$ and $0.3$ for $\rho_B= 2/3$ and $\rho_B= 5/6$, respectively. Comparing Figs. ~\ref{fig3} and ~\ref{fig8}, we note that the attractive interactions between the fermions prohibits the mixed Mott insulator state and facilitates the emergence of the spin-selective insulators.\par
\section{\label{sec5} Conclusions}
In the present work we showed that a spin imbalance in mixtures of fermionic and bosonic atoms leads to the emergence of fully-polarized insulating phases. For this, the effect of spin population imbalance $I=(\hat{N}_F^\downarrow-\hat{N}_F^\uparrow)/
(\hat{N}_F^\uparrow+\hat{N}_F^\downarrow)$ on superfluid-insulator transitions was explored in mixtures composed of scalar bosons and two-color fermions in the hard-core aproximation. Using the density matrix renormalization group method, we calculated the bosonic chemical potential as a function of the number of bosons and the interaction parameters, building several zero-temperature phase diagrams.\par 
The number of incompressible states in a balanced mixture depends on the repulsive or attractive character of the fermion-fermion interaction. This is, a noncommensurate state is always present regardless of the sign of $U_{FF}$, while the mixed Mott state only emerges for a repulsive interaction \cite{Avella-PRA20}. For any spin population asymmetry and repulsive interaction between fermions, we found that only one noncommensurate plateau is affected by the imbalance, leaving the other unaltered regardless of the sign of the boson-fermion coupling. The noncommensurate insulator state is divided into two spin-selective insulator states, which are separated by a superfluid phase.\par 
We obtained that for a repulsive boson-fermion interaction, the new incompressible states fulfill the relations $\rho_B+\frac{1}{2}\rho_F(1 \pm I)=1$, in terms of the global densities $\rho_{B,(F)}$ of the carriers and the imbalance $I$. The above insulator states are composed by a gapless state for one kind of fermions and an  insulator state for the other, with the latter satisfying a commensurability relation with the bosons. Namely, in a gapless fermion polarized background, we have $\rho_B+\rho^{\uparrow,(\downarrow)}_F=1$ for the spin-selective states, indicating that the number of bosons plus the number of one kind of fermions is commensurate with the lattice size in each case. Special distributions of carriers were observed in fermionic half-filling, which are signaled by a maximum of the charge structure factor. In particular, for an imbalance of $I=1/2$, we obtained a dimer distribution of carriers for both insulator states.\par  
On the other hand, attractive boson-fermion interactions and imbalance generate two new insulator states that fulfill $\rho_B+\frac{1}{2}\rho_F(-1 \pm I)=0$, from which we established that each spin-selective state satisfies 
$\rho_B-\rho^{\uparrow,(\downarrow)}_F=0$, in a gapless fermion polarized background. Note that the results for repulsive and attractive boson-fermion coupling are related by $| \mu^{A}_{B}\left(\rho_B\right)|=| \mu^{R}_{B}\left(1-\rho_B\right)|$.\par
We expect that our results motivate further research on Bose-Fermi mixtures where a spin imbalance plays a key role, such as certain open fermionic systems~\cite{Buca-NP19,Buca-PRL19}. In addition, we believe that the experimental observation of the spin-selective insulators suggested in the present investigation is an intriguing challenge, which can be faced with state-of-the-art cold-atom setups.\par 
\appendix*
\section{\label{sec6} Correlations}
\begin{figure}[t]
	\centering
	\includegraphics[width=21pc]{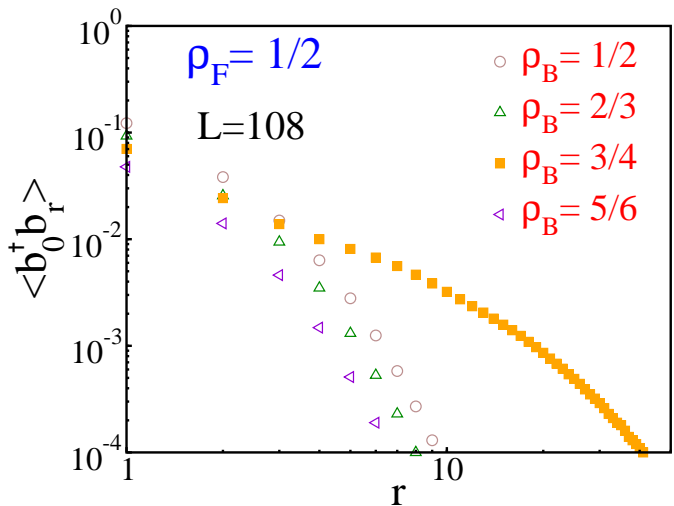}
	\caption{Bosonic correlations $\langle b^{\dagger}_0b_r\rangle$ as a function of $r$ for a fermionic density $\rho_F=1/2$, imbalance $I=1/3$, repulsive fermion-fermion interaction $U_{FF}=6$ and repulsive inter-species interaction $U_{BF}=10$. Open circle brown, green triangle up, violet triangle left and solid orange square dots correspond to the bosonic densities $\rho_B= 1/2, 2/3, 5/6$, and $3/4$, respectively.}
	\label{fig9}
\end{figure}

In the main text, we explored the ground-state of a Bose-Fermi mixture composed of two-color fermions and hard-core bosons. We found and characterized gapless and gapped states, and expressed their location in terms of closed-relations.
However, the former ones need a more demanding exploration, and we denominated these regions surrounding the insulator states as superfluid following the literature on this subject. To show the emergence of a superfluid state for some bosonic densities due to the imbalance, we calculate the correlation function $\langle b^{\dagger}_0b_r\rangle$ as a function of $r$ for different bosonic densities. In addition, we keep the fermionic density fixed at $\rho_F=1/2$, and use the imbalance and interaction parameters of Fig.~\ref{fig2} (a), obtaining the results shown in Fig.~\ref{fig9}.\par
Considering a balanced Bose-Fermi mixture with a fermionic density  $\rho_F=1/2$ and repulsive interaction parameters, an insulator state emerges at the bosonic density  $\rho_B=3/4$. However turning on the imbalance leads to the splitting of this insulator and to a gapless state for this particular bosonic density. In Fig.~\ref{fig9}, we show the evolution of the bosonic correlation $\langle b^{\dagger}_0b_r\rangle$ as a function of $r$ at the bosonic density $\rho_B=3/4$, and we observe that the curve is similar to that for bare bosons ~\cite{Kuhner-PRB00}, which can be fitted by  $\langle b^{\dagger}_0b_r\rangle \sim \lvert r \rvert^{-K^{*}/2}$. In fact, several fittings considering different ranges of $r$ gave values of $K^{*}>2$. This suggests that the ground state for this bosonic density will be a polaronic Luttinger liquid with $K^{*}>2$ ~\cite{Mathey-PRL04}. However more calculations are necessary to determine a precise value of $K^{*}$, which is beyond the scope of this article as it focuses on establishing the conditions for the emergence of insulators (note that the characterization of the FFLO superfluid was discussed by Singh and Orso~\cite{Singh-PRR20}). Also, in Fig.~\ref{fig9} we see that the bosonic correlations exhibit the same behavior for the bosonic densities $\rho_B=1/2,2/3,5/6$, for which an exponential decay is evidenced; this reinforces the fact that for these densities we have insulating states. On the other hand, for $\rho_B=3/4$, the correlations fulfill a power-law due to the imbalance, establishing that the spin-selective insulators are separated by a gapless Luttinger liquid state.
\section*{Acknowledgments}
R. G-S thanks the support of the scholarship ``Grado de Honor'' of Universidad Nacional de Colombia. J. J. M-A thanks the support of Ministerio de Ciencia, Tecnolog\'{\i}a e Innovaci\'on (MINCIENCIAS), through the project \textit{Producci\'on y Caracterizaci\'on de Nuevos Materiales Cu\'anticos de Baja Dimensionalidad: Criticalidad Cu\'antica y Transiciones de Fase Electr\'onicas} (Grant No. 120480863414).

\bibliography{Bibliografia}

\begin{thebibliography}{86}%
\makeatletter
\providecommand \@ifxundefined [1]{%
 \@ifx{#1\undefined}
}%
\providecommand \@ifnum [1]{%
 \ifnum #1\expandafter \@firstoftwo
 \else \expandafter \@secondoftwo
 \fi
}%
\providecommand \@ifx [1]{%
 \ifx #1\expandafter \@firstoftwo
 \else \expandafter \@secondoftwo
 \fi
}%
\providecommand \natexlab [1]{#1}%
\providecommand \enquote  [1]{``#1''}%
\providecommand \bibnamefont  [1]{#1}%
\providecommand \bibfnamefont [1]{#1}%
\providecommand \citenamefont [1]{#1}%
\providecommand \href@noop [0]{\@secondoftwo}%
\providecommand \href [0]{\begingroup \@sanitize@url \@href}%
\providecommand \@href[1]{\@@startlink{#1}\@@href}%
\providecommand \@@href[1]{\endgroup#1\@@endlink}%
\providecommand \@sanitize@url [0]{\catcode `\\12\catcode `\$12\catcode
  `\&12\catcode `\#12\catcode `\^12\catcode `\_12\catcode `\%12\relax}%
\providecommand \@@startlink[1]{}%
\providecommand \@@endlink[0]{}%
\providecommand \url  [0]{\begingroup\@sanitize@url \@url }%
\providecommand \@url [1]{\endgroup\@href {#1}{\urlprefix }}%
\providecommand \urlprefix  [0]{URL }%
\providecommand \Eprint [0]{\href }%
\providecommand \doibase [0]{http://dx.doi.org/}%
\providecommand \selectlanguage [0]{\@gobble}%
\providecommand \bibinfo  [0]{\@secondoftwo}%
\providecommand \bibfield  [0]{\@secondoftwo}%
\providecommand \translation [1]{[#1]}%
\providecommand \BibitemOpen [0]{}%
\providecommand \bibitemStop [0]{}%
\providecommand \bibitemNoStop [0]{.\EOS\space}%
\providecommand \EOS [0]{\spacefactor3000\relax}%
\providecommand \BibitemShut  [1]{\csname bibitem#1\endcsname}%
\let\auto@bib@innerbib\@empty
\bibitem [{\citenamefont {Bloch}\ \emph {et~al.}(2008)\citenamefont {Bloch},
  \citenamefont {Dalibard},\ and\ \citenamefont {Zwerger}}]{IBloch-RMP08}%
  \BibitemOpen
  \bibfield  {author} {\bibinfo {author} {\bibfnamefont {I.}~\bibnamefont
  {Bloch}}, \bibinfo {author} {\bibfnamefont {J.}~\bibnamefont {Dalibard}}, \
  and\ \bibinfo {author} {\bibfnamefont {W.}~\bibnamefont {Zwerger}},\
  }\href@noop {} {\bibfield  {journal} {\bibinfo  {journal} {Rev. Mod. Phys.}\
  }\textbf {\bibinfo {volume} {80}},\ \bibinfo {pages} {885} (\bibinfo {year}
  {2008})}\BibitemShut {NoStop}%
\bibitem [{\citenamefont {Esslinger}(2010)}]{Esslinger-AR10}%
  \BibitemOpen
  \bibfield  {author} {\bibinfo {author} {\bibfnamefont {T.}~\bibnamefont
  {Esslinger}},\ }\href@noop {} {\bibfield  {journal} {\bibinfo  {journal}
  {Annu. Rev. Condens. Matter Phys.}\ }\textbf {\bibinfo {volume} {1}},\
  \bibinfo {pages} {129} (\bibinfo {year} {2010})}\BibitemShut {NoStop}%
\bibitem [{\citenamefont {Bloch}\ \emph {et~al.}(2012)\citenamefont {Bloch},
  \citenamefont {Dalibard},\ and\ \citenamefont
  {Nascimb{\'e}ne}}]{IBloch-NP12}%
  \BibitemOpen
  \bibfield  {author} {\bibinfo {author} {\bibfnamefont {I.}~\bibnamefont
  {Bloch}}, \bibinfo {author} {\bibfnamefont {J.}~\bibnamefont {Dalibard}}, \
  and\ \bibinfo {author} {\bibfnamefont {S.}~\bibnamefont {Nascimb{\'e}ne}},\
  }\href@noop {} {\bibfield  {journal} {\bibinfo  {journal} {Nat. Phys.}\
  }\textbf {\bibinfo {volume} {8}},\ \bibinfo {pages} {267} (\bibinfo {year}
  {2012})}\BibitemShut {NoStop}%
\bibitem [{\citenamefont {Gross}\ and\ \citenamefont
  {Bloch}(2017)}]{Gross-S17}%
  \BibitemOpen
  \bibfield  {author} {\bibinfo {author} {\bibfnamefont {C.}~\bibnamefont
  {Gross}}\ and\ \bibinfo {author} {\bibfnamefont {I.}~\bibnamefont {Bloch}},\
  }\href@noop {} {\bibfield  {journal} {\bibinfo  {journal} {Science}\ }\textbf
  {\bibinfo {volume} {357}},\ \bibinfo {pages} {995} (\bibinfo {year}
  {2017})}\BibitemShut {NoStop}%
\bibitem [{\citenamefont {Greiner}\ \emph {et~al.}(2002)\citenamefont
  {Greiner}, \citenamefont {Mandel}, \citenamefont {Esslinger}, \citenamefont
  {H{\"a}nsch},\ and\ \citenamefont {Bloch}}]{Greiner-N02a}%
  \BibitemOpen
  \bibfield  {author} {\bibinfo {author} {\bibfnamefont {M.}~\bibnamefont
  {Greiner}}, \bibinfo {author} {\bibfnamefont {O.}~\bibnamefont {Mandel}},
  \bibinfo {author} {\bibfnamefont {T.}~\bibnamefont {Esslinger}}, \bibinfo
  {author} {\bibfnamefont {T.~W.}\ \bibnamefont {H{\"a}nsch}}, \ and\ \bibinfo
  {author} {\bibfnamefont {I.}~\bibnamefont {Bloch}},\ }\href@noop {}
  {\bibfield  {journal} {\bibinfo  {journal} {Nature}\ }\textbf {\bibinfo
  {volume} {415}},\ \bibinfo {pages} {39} (\bibinfo {year} {2002})}\BibitemShut
  {NoStop}%
\bibitem [{\citenamefont {J{\"o}rdens}\ \emph {et~al.}(2008)\citenamefont
  {J{\"o}rdens}, \citenamefont {Strohmaier}, \citenamefont {G{\"u}nter},
  \citenamefont {Moritz},\ and\ \citenamefont {Esslinger}}]{Jordens-N08}%
  \BibitemOpen
  \bibfield  {author} {\bibinfo {author} {\bibfnamefont {R.}~\bibnamefont
  {J{\"o}rdens}}, \bibinfo {author} {\bibfnamefont {N.}~\bibnamefont
  {Strohmaier}}, \bibinfo {author} {\bibfnamefont {K.}~\bibnamefont
  {G{\"u}nter}}, \bibinfo {author} {\bibfnamefont {H.}~\bibnamefont {Moritz}},
  \ and\ \bibinfo {author} {\bibfnamefont {T.}~\bibnamefont {Esslinger}},\
  }\href@noop {} {\bibfield  {journal} {\bibinfo  {journal} {Nature}\ }\textbf
  {\bibinfo {volume} {455}},\ \bibinfo {pages} {204} (\bibinfo {year}
  {2008})}\BibitemShut {NoStop}%
\bibitem [{\citenamefont {Schneider}\ \emph {et~al.}(2008)\citenamefont
  {Schneider}, \citenamefont {Hackerm{\"u}ller}, \citenamefont {Will},
  \citenamefont {Best}, \citenamefont {Bloch}, \citenamefont {Costi},
  \citenamefont {Helmes}, \citenamefont {Rasch},\ and\ \citenamefont
  {Rosch}}]{Schneider-S08}%
  \BibitemOpen
  \bibfield  {author} {\bibinfo {author} {\bibfnamefont {U.}~\bibnamefont
  {Schneider}}, \bibinfo {author} {\bibfnamefont {L.}~\bibnamefont
  {Hackerm{\"u}ller}}, \bibinfo {author} {\bibfnamefont {S.}~\bibnamefont
  {Will}}, \bibinfo {author} {\bibfnamefont {T.}~\bibnamefont {Best}}, \bibinfo
  {author} {\bibfnamefont {I.}~\bibnamefont {Bloch}}, \bibinfo {author}
  {\bibfnamefont {T.~A.}\ \bibnamefont {Costi}}, \bibinfo {author}
  {\bibfnamefont {R.~W.}\ \bibnamefont {Helmes}}, \bibinfo {author}
  {\bibfnamefont {D.}~\bibnamefont {Rasch}}, \ and\ \bibinfo {author}
  {\bibfnamefont {A.}~\bibnamefont {Rosch}},\ }\href@noop {} {\bibfield
  {journal} {\bibinfo  {journal} {Science}\ }\textbf {\bibinfo {volume}
  {322}},\ \bibinfo {pages} {1520} (\bibinfo {year} {2008})}\BibitemShut
  {NoStop}%
\bibitem [{\citenamefont {Bakr}\ \emph {et~al.}(2010)\citenamefont {Bakr},
  \citenamefont {Peng}, \citenamefont {Tai}, \citenamefont {Ma}, \citenamefont
  {Simon}, \citenamefont {Gillen}, \citenamefont {F{\"o}lling}, \citenamefont
  {Pollet},\ and\ \citenamefont {Greiner}}]{Bakr-S10}%
  \BibitemOpen
  \bibfield  {author} {\bibinfo {author} {\bibfnamefont {W.~S.}\ \bibnamefont
  {Bakr}}, \bibinfo {author} {\bibfnamefont {A.}~\bibnamefont {Peng}}, \bibinfo
  {author} {\bibfnamefont {M.~E.}\ \bibnamefont {Tai}}, \bibinfo {author}
  {\bibfnamefont {R.}~\bibnamefont {Ma}}, \bibinfo {author} {\bibfnamefont
  {J.}~\bibnamefont {Simon}}, \bibinfo {author} {\bibfnamefont {J.~I.}\
  \bibnamefont {Gillen}}, \bibinfo {author} {\bibfnamefont {S.}~\bibnamefont
  {F{\"o}lling}}, \bibinfo {author} {\bibfnamefont {L.}~\bibnamefont {Pollet}},
  \ and\ \bibinfo {author} {\bibfnamefont {M.}~\bibnamefont {Greiner}},\
  }\href@noop {} {\bibfield  {journal} {\bibinfo  {journal} {Science}\ }\textbf
  {\bibinfo {volume} {547}},\ \bibinfo {pages} {329} (\bibinfo {year}
  {2010})}\BibitemShut {NoStop}%
\bibitem [{\citenamefont {Sherson}\ \emph {et~al.}(2010)\citenamefont
  {Sherson}, \citenamefont {Weitenberg}, \citenamefont {Endres}, \citenamefont
  {Cheneau}, \citenamefont {Bloch},\ and\ \citenamefont
  {Kuhr}}]{Sherson-Nat10}%
  \BibitemOpen
  \bibfield  {author} {\bibinfo {author} {\bibfnamefont {J.~F.}\ \bibnamefont
  {Sherson}}, \bibinfo {author} {\bibfnamefont {C.}~\bibnamefont {Weitenberg}},
  \bibinfo {author} {\bibfnamefont {M.}~\bibnamefont {Endres}}, \bibinfo
  {author} {\bibfnamefont {M.}~\bibnamefont {Cheneau}}, \bibinfo {author}
  {\bibfnamefont {I.}~\bibnamefont {Bloch}}, \ and\ \bibinfo {author}
  {\bibfnamefont {S.}~\bibnamefont {Kuhr}},\ }\href@noop {} {\bibfield
  {journal} {\bibinfo  {journal} {Nature}\ }\textbf {\bibinfo {volume} {467}},\
  \bibinfo {pages} {68} (\bibinfo {year} {2010})}\BibitemShut {NoStop}%
\bibitem [{\citenamefont {Greif}\ \emph {et~al.}(2016)\citenamefont {Greif},
  \citenamefont {Parsons}, \citenamefont {Mazurenko}, \citenamefont {Chiu},
  \citenamefont {Blatt}, \citenamefont {Huber}, \citenamefont {Ji},\ and\
  \citenamefont {Greiner}}]{Greif-Science16}%
  \BibitemOpen
  \bibfield  {author} {\bibinfo {author} {\bibfnamefont {D.}~\bibnamefont
  {Greif}}, \bibinfo {author} {\bibfnamefont {M.~F.}\ \bibnamefont {Parsons}},
  \bibinfo {author} {\bibfnamefont {A.}~\bibnamefont {Mazurenko}}, \bibinfo
  {author} {\bibfnamefont {C.~S.}\ \bibnamefont {Chiu}}, \bibinfo {author}
  {\bibfnamefont {S.}~\bibnamefont {Blatt}}, \bibinfo {author} {\bibfnamefont
  {F.}~\bibnamefont {Huber}}, \bibinfo {author} {\bibfnamefont
  {G.}~\bibnamefont {Ji}}, \ and\ \bibinfo {author} {\bibfnamefont
  {M.}~\bibnamefont {Greiner}},\ }\href@noop {} {\bibfield  {journal} {\bibinfo
   {journal} {Science}\ }\textbf {\bibinfo {volume} {351}},\ \bibinfo {pages}
  {953} (\bibinfo {year} {2016})}\BibitemShut {NoStop}%
\bibitem [{\citenamefont {Truscott}\ \emph {et~al.}(2001)\citenamefont
  {Truscott}, \citenamefont {Strecker}, \citenamefont {McAlexander},
  \citenamefont {Partridge},\ and\ \citenamefont {Hulet}}]{Truscott-S01}%
  \BibitemOpen
  \bibfield  {author} {\bibinfo {author} {\bibfnamefont {A.~G.}\ \bibnamefont
  {Truscott}}, \bibinfo {author} {\bibfnamefont {K.~E.}\ \bibnamefont
  {Strecker}}, \bibinfo {author} {\bibfnamefont {W.~I.}\ \bibnamefont
  {McAlexander}}, \bibinfo {author} {\bibfnamefont {G.~B.}\ \bibnamefont
  {Partridge}}, \ and\ \bibinfo {author} {\bibfnamefont {R.~G.}\ \bibnamefont
  {Hulet}},\ }\href@noop {} {\bibfield  {journal} {\bibinfo  {journal}
  {Science}\ }\textbf {\bibinfo {volume} {291}},\ \bibinfo {pages} {2570}
  (\bibinfo {year} {2001})}\BibitemShut {NoStop}%
\bibitem [{\citenamefont {Schreck}\ \emph {et~al.}(2001)\citenamefont
  {Schreck}, \citenamefont {Khaykovich}, \citenamefont {Corwin}, \citenamefont
  {Ferrari}, \citenamefont {Bourdel}, \citenamefont {Cubizolles},\ and\
  \citenamefont {Salomon}}]{Schreck-PRL01}%
  \BibitemOpen
  \bibfield  {author} {\bibinfo {author} {\bibfnamefont {F.}~\bibnamefont
  {Schreck}}, \bibinfo {author} {\bibfnamefont {L.}~\bibnamefont {Khaykovich}},
  \bibinfo {author} {\bibfnamefont {K.~L.}\ \bibnamefont {Corwin}}, \bibinfo
  {author} {\bibfnamefont {G.}~\bibnamefont {Ferrari}}, \bibinfo {author}
  {\bibfnamefont {T.}~\bibnamefont {Bourdel}}, \bibinfo {author} {\bibfnamefont
  {J.}~\bibnamefont {Cubizolles}}, \ and\ \bibinfo {author} {\bibfnamefont
  {C.}~\bibnamefont {Salomon}},\ }\href@noop {} {\bibfield  {journal} {\bibinfo
   {journal} {Phys. Rev. Lett.}\ }\textbf {\bibinfo {volume} {87}},\ \bibinfo
  {pages} {080403} (\bibinfo {year} {2001})}\BibitemShut {NoStop}%
\bibitem [{\citenamefont {Hadzibabic}\ \emph {et~al.}(2002)\citenamefont
  {Hadzibabic}, \citenamefont {Stan}, \citenamefont {Dieckmann}, \citenamefont
  {Gupta}, \citenamefont {Zwierlein}, \citenamefont {G{\"o}rlitz},\ and\
  \citenamefont {Ketterle}}]{Hadzibabic-PRL02}%
  \BibitemOpen
  \bibfield  {author} {\bibinfo {author} {\bibfnamefont {Z.}~\bibnamefont
  {Hadzibabic}}, \bibinfo {author} {\bibfnamefont {C.~A.}\ \bibnamefont
  {Stan}}, \bibinfo {author} {\bibfnamefont {K.}~\bibnamefont {Dieckmann}},
  \bibinfo {author} {\bibfnamefont {S.}~\bibnamefont {Gupta}}, \bibinfo
  {author} {\bibfnamefont {M.~W.}\ \bibnamefont {Zwierlein}}, \bibinfo {author}
  {\bibfnamefont {A.}~\bibnamefont {G{\"o}rlitz}}, \ and\ \bibinfo {author}
  {\bibfnamefont {W.}~\bibnamefont {Ketterle}},\ }\href@noop {} {\bibfield
  {journal} {\bibinfo  {journal} {Phys. Rev. Lett.}\ }\textbf {\bibinfo
  {volume} {88}},\ \bibinfo {pages} {160401} (\bibinfo {year}
  {2002})}\BibitemShut {NoStop}%
\bibitem [{\citenamefont {Roati}\ \emph {et~al.}(2002)\citenamefont {Roati},
  \citenamefont {Riboli}, \citenamefont {Modugno},\ and\ \citenamefont
  {Inguscio}}]{Roati-PRL02}%
  \BibitemOpen
  \bibfield  {author} {\bibinfo {author} {\bibfnamefont {G.}~\bibnamefont
  {Roati}}, \bibinfo {author} {\bibfnamefont {F.}~\bibnamefont {Riboli}},
  \bibinfo {author} {\bibfnamefont {G.}~\bibnamefont {Modugno}}, \ and\
  \bibinfo {author} {\bibfnamefont {M.}~\bibnamefont {Inguscio}},\ }\href@noop
  {} {\bibfield  {journal} {\bibinfo  {journal} {Phys. Rev. Lett.}\ }\textbf
  {\bibinfo {volume} {89}},\ \bibinfo {pages} {150403} (\bibinfo {year}
  {2002})}\BibitemShut {NoStop}%
\bibitem [{\citenamefont {Ott}\ \emph {et~al.}(2004)\citenamefont {Ott},
  \citenamefont {de~Mirandes}, \citenamefont {Ferlaino}, \citenamefont {Roati},
  \citenamefont {Modugno},\ and\ \citenamefont {Inguscio}}]{Ott-PRL04}%
  \BibitemOpen
  \bibfield  {author} {\bibinfo {author} {\bibfnamefont {H.}~\bibnamefont
  {Ott}}, \bibinfo {author} {\bibfnamefont {E.}~\bibnamefont {de~Mirandes}},
  \bibinfo {author} {\bibfnamefont {F.}~\bibnamefont {Ferlaino}}, \bibinfo
  {author} {\bibfnamefont {G.}~\bibnamefont {Roati}}, \bibinfo {author}
  {\bibfnamefont {G.}~\bibnamefont {Modugno}}, \ and\ \bibinfo {author}
  {\bibfnamefont {M.}~\bibnamefont {Inguscio}},\ }\href@noop {} {\bibfield
  {journal} {\bibinfo  {journal} {Phys. Rev. Lett.}\ }\textbf {\bibinfo
  {volume} {92}},\ \bibinfo {pages} {160601} (\bibinfo {year}
  {2004})}\BibitemShut {NoStop}%
\bibitem [{\citenamefont {Silber}\ \emph {et~al.}(2005)\citenamefont {Silber},
  \citenamefont {G{\"u}nther}, \citenamefont {Marzok}, \citenamefont {Deh},
  \citenamefont {Courteille},\ and\ \citenamefont {Zimmermann}}]{Silber-PRL05}%
  \BibitemOpen
  \bibfield  {author} {\bibinfo {author} {\bibfnamefont {C.}~\bibnamefont
  {Silber}}, \bibinfo {author} {\bibfnamefont {S.}~\bibnamefont {G{\"u}nther}},
  \bibinfo {author} {\bibfnamefont {C.}~\bibnamefont {Marzok}}, \bibinfo
  {author} {\bibfnamefont {B.}~\bibnamefont {Deh}}, \bibinfo {author}
  {\bibfnamefont {P.~W.}\ \bibnamefont {Courteille}}, \ and\ \bibinfo {author}
  {\bibfnamefont {C.}~\bibnamefont {Zimmermann}},\ }\href@noop {} {\bibfield
  {journal} {\bibinfo  {journal} {Phys. Rev. Lett.}\ }\textbf {\bibinfo
  {volume} {95}},\ \bibinfo {pages} {170408} (\bibinfo {year}
  {2005})}\BibitemShut {NoStop}%
\bibitem [{\citenamefont {G{\"u}nter}\ \emph {et~al.}(2006)\citenamefont
  {G{\"u}nter}, \citenamefont {St{\"o}ferle}, \citenamefont {Moritz},
  \citenamefont {K{\"o}hl},\ and\ \citenamefont {Esslinger}}]{Gunter-PRL06}%
  \BibitemOpen
  \bibfield  {author} {\bibinfo {author} {\bibfnamefont {K.}~\bibnamefont
  {G{\"u}nter}}, \bibinfo {author} {\bibfnamefont {T.}~\bibnamefont
  {St{\"o}ferle}}, \bibinfo {author} {\bibfnamefont {H.}~\bibnamefont
  {Moritz}}, \bibinfo {author} {\bibfnamefont {M.}~\bibnamefont {K{\"o}hl}}, \
  and\ \bibinfo {author} {\bibfnamefont {T.}~\bibnamefont {Esslinger}},\
  }\href@noop {} {\bibfield  {journal} {\bibinfo  {journal} {Phys. Rev. Lett.}\
  }\textbf {\bibinfo {volume} {96}},\ \bibinfo {pages} {180402} (\bibinfo
  {year} {2006})}\BibitemShut {NoStop}%
\bibitem [{\citenamefont {Ospelkaus}\ \emph {et~al.}(2006)\citenamefont
  {Ospelkaus}, \citenamefont {Ospelkaus}, \citenamefont {Wille}, \citenamefont
  {Succo}, \citenamefont {Ernst}, \citenamefont {Sengstock},\ and\
  \citenamefont {Bongs}}]{Ospelkaus-PRL06}%
  \BibitemOpen
  \bibfield  {author} {\bibinfo {author} {\bibfnamefont {S.}~\bibnamefont
  {Ospelkaus}}, \bibinfo {author} {\bibfnamefont {C.}~\bibnamefont
  {Ospelkaus}}, \bibinfo {author} {\bibfnamefont {O.}~\bibnamefont {Wille}},
  \bibinfo {author} {\bibfnamefont {M.}~\bibnamefont {Succo}}, \bibinfo
  {author} {\bibfnamefont {P.}~\bibnamefont {Ernst}}, \bibinfo {author}
  {\bibfnamefont {K.}~\bibnamefont {Sengstock}}, \ and\ \bibinfo {author}
  {\bibfnamefont {K.}~\bibnamefont {Bongs}},\ }\href@noop {} {\bibfield
  {journal} {\bibinfo  {journal} {Phys. Rev. Lett.}\ }\textbf {\bibinfo
  {volume} {96}},\ \bibinfo {pages} {180403} (\bibinfo {year}
  {2006})}\BibitemShut {NoStop}%
\bibitem [{\citenamefont {Zaccanti}\ \emph {et~al.}(2006)\citenamefont
  {Zaccanti}, \citenamefont {D’Errico}, \citenamefont {Ferlaino},
  \citenamefont {Roati}, \citenamefont {Inguscio},\ and\ \citenamefont
  {Modugno}}]{Zaccanti-PRA06}%
  \BibitemOpen
  \bibfield  {author} {\bibinfo {author} {\bibfnamefont {M.}~\bibnamefont
  {Zaccanti}}, \bibinfo {author} {\bibfnamefont {C.}~\bibnamefont
  {D’Errico}}, \bibinfo {author} {\bibfnamefont {F.}~\bibnamefont
  {Ferlaino}}, \bibinfo {author} {\bibfnamefont {G.}~\bibnamefont {Roati}},
  \bibinfo {author} {\bibfnamefont {M.}~\bibnamefont {Inguscio}}, \ and\
  \bibinfo {author} {\bibfnamefont {G.}~\bibnamefont {Modugno}},\ }\href@noop
  {} {\bibfield  {journal} {\bibinfo  {journal} {Phys. Rev. A}\ }\textbf
  {\bibinfo {volume} {74}},\ \bibinfo {pages} {041605(R)} (\bibinfo {year}
  {2006})}\BibitemShut {NoStop}%
\bibitem [{\citenamefont {McNamara}\ \emph {et~al.}(2006)\citenamefont
  {McNamara}, \citenamefont {Jeltes}, \citenamefont {Tychkov}, \citenamefont
  {Hogervorst},\ and\ \citenamefont {Vassen}}]{McNamara-PRL06}%
  \BibitemOpen
  \bibfield  {author} {\bibinfo {author} {\bibfnamefont {J.~M.}\ \bibnamefont
  {McNamara}}, \bibinfo {author} {\bibfnamefont {T.}~\bibnamefont {Jeltes}},
  \bibinfo {author} {\bibfnamefont {A.~S.}\ \bibnamefont {Tychkov}}, \bibinfo
  {author} {\bibfnamefont {W.}~\bibnamefont {Hogervorst}}, \ and\ \bibinfo
  {author} {\bibfnamefont {W.}~\bibnamefont {Vassen}},\ }\href@noop {}
  {\bibfield  {journal} {\bibinfo  {journal} {Phys. Rev. Lett.}\ }\textbf
  {\bibinfo {volume} {97}},\ \bibinfo {pages} {080404} (\bibinfo {year}
  {2006})}\BibitemShut {NoStop}%
\bibitem [{\citenamefont {Best}\ \emph {et~al.}(2009)\citenamefont {Best},
  \citenamefont {Will}, \citenamefont {Schneider}, \citenamefont
  {Hackerm{\"u}ller}, \citenamefont {van Oosten}, \citenamefont {Bloch},\ and\
  \citenamefont {L{\"u}hmann}}]{Best-PRL09}%
  \BibitemOpen
  \bibfield  {author} {\bibinfo {author} {\bibfnamefont {T.}~\bibnamefont
  {Best}}, \bibinfo {author} {\bibfnamefont {S.}~\bibnamefont {Will}}, \bibinfo
  {author} {\bibfnamefont {U.}~\bibnamefont {Schneider}}, \bibinfo {author}
  {\bibfnamefont {L.}~\bibnamefont {Hackerm{\"u}ller}}, \bibinfo {author}
  {\bibfnamefont {D.}~\bibnamefont {van Oosten}}, \bibinfo {author}
  {\bibfnamefont {I.}~\bibnamefont {Bloch}}, \ and\ \bibinfo {author}
  {\bibfnamefont {D.-S.}\ \bibnamefont {L{\"u}hmann}},\ }\href@noop {}
  {\bibfield  {journal} {\bibinfo  {journal} {Phys. Rev. Lett.}\ }\textbf
  {\bibinfo {volume} {102}},\ \bibinfo {pages} {030408} (\bibinfo {year}
  {2009})}\BibitemShut {NoStop}%
\bibitem [{\citenamefont {Fukuhara}\ \emph {et~al.}(2009)\citenamefont
  {Fukuhara}, \citenamefont {Sugawa}, \citenamefont {Takasu},\ and\
  \citenamefont {Takahashi}}]{Fukuhara-PRA09b}%
  \BibitemOpen
  \bibfield  {author} {\bibinfo {author} {\bibfnamefont {T.}~\bibnamefont
  {Fukuhara}}, \bibinfo {author} {\bibfnamefont {S.}~\bibnamefont {Sugawa}},
  \bibinfo {author} {\bibfnamefont {Y.}~\bibnamefont {Takasu}}, \ and\ \bibinfo
  {author} {\bibfnamefont {Y.}~\bibnamefont {Takahashi}},\ }\href@noop {}
  {\bibfield  {journal} {\bibinfo  {journal} {Phys. Rev. A}\ }\textbf {\bibinfo
  {volume} {79}},\ \bibinfo {pages} {021601(R)} (\bibinfo {year}
  {2009})}\BibitemShut {NoStop}%
\bibitem [{\citenamefont {Deh}\ \emph {et~al.}(2010)\citenamefont {Deh},
  \citenamefont {Gunton}, \citenamefont {Klappauf}, \citenamefont {Li},
  \citenamefont {Semczuk}, \citenamefont {Dongen},\ and\ \citenamefont
  {Madison}}]{Deh-PRA10}%
  \BibitemOpen
  \bibfield  {author} {\bibinfo {author} {\bibfnamefont {B.}~\bibnamefont
  {Deh}}, \bibinfo {author} {\bibfnamefont {W.}~\bibnamefont {Gunton}},
  \bibinfo {author} {\bibfnamefont {B.~G.}\ \bibnamefont {Klappauf}}, \bibinfo
  {author} {\bibfnamefont {Z.}~\bibnamefont {Li}}, \bibinfo {author}
  {\bibfnamefont {M.}~\bibnamefont {Semczuk}}, \bibinfo {author} {\bibfnamefont
  {J.~V.}\ \bibnamefont {Dongen}}, \ and\ \bibinfo {author} {\bibfnamefont
  {K.~W.}\ \bibnamefont {Madison}},\ }\href@noop {} {\bibfield  {journal}
  {\bibinfo  {journal} {Phys. Rev. A}\ }\textbf {\bibinfo {volume} {82}},\
  \bibinfo {pages} {020701(R)} (\bibinfo {year} {2010})}\BibitemShut {NoStop}%
\bibitem [{\citenamefont {Tey}\ \emph {et~al.}(2010)\citenamefont {Tey},
  \citenamefont {Stellmer}, \citenamefont {Grimm},\ and\ \citenamefont
  {Schreck}}]{Tey-PRA10}%
  \BibitemOpen
  \bibfield  {author} {\bibinfo {author} {\bibfnamefont {M.~K.}\ \bibnamefont
  {Tey}}, \bibinfo {author} {\bibfnamefont {S.}~\bibnamefont {Stellmer}},
  \bibinfo {author} {\bibfnamefont {R.}~\bibnamefont {Grimm}}, \ and\ \bibinfo
  {author} {\bibfnamefont {F.}~\bibnamefont {Schreck}},\ }\href@noop {}
  {\bibfield  {journal} {\bibinfo  {journal} {Phys. Rev. A}\ }\textbf {\bibinfo
  {volume} {82}},\ \bibinfo {pages} {011608(R)} (\bibinfo {year}
  {2010})}\BibitemShut {NoStop}%
\bibitem [{\citenamefont {Sugawa}\ \emph {et~al.}(2011)\citenamefont {Sugawa},
  \citenamefont {Inaba}, \citenamefont {Taie}, \citenamefont {Yamazaki},
  \citenamefont {Yamashita},\ and\ \citenamefont {Takahashi}}]{Sugawa-NP11}%
  \BibitemOpen
  \bibfield  {author} {\bibinfo {author} {\bibfnamefont {S.}~\bibnamefont
  {Sugawa}}, \bibinfo {author} {\bibfnamefont {K.}~\bibnamefont {Inaba}},
  \bibinfo {author} {\bibfnamefont {S.}~\bibnamefont {Taie}}, \bibinfo {author}
  {\bibfnamefont {R.}~\bibnamefont {Yamazaki}}, \bibinfo {author}
  {\bibfnamefont {M.}~\bibnamefont {Yamashita}}, \ and\ \bibinfo {author}
  {\bibfnamefont {Y.}~\bibnamefont {Takahashi}},\ }\href@noop {} {\bibfield
  {journal} {\bibinfo  {journal} {Nat. Phys.}\ }\textbf {\bibinfo {volume}
  {7}},\ \bibinfo {pages} {642} (\bibinfo {year} {2011})}\BibitemShut {NoStop}%
\bibitem [{\citenamefont {Schuster}\ \emph {et~al.}(2012)\citenamefont
  {Schuster}, \citenamefont {Scelle}, \citenamefont {Trautmann}, \citenamefont
  {Knoop}, \citenamefont {Oberthaler}, \citenamefont {Haverhals}, \citenamefont
  {Goosen}, \citenamefont {Kokkelmans},\ and\ \citenamefont
  {Tiemann}}]{Schuster-PRA12}%
  \BibitemOpen
  \bibfield  {author} {\bibinfo {author} {\bibfnamefont {T.}~\bibnamefont
  {Schuster}}, \bibinfo {author} {\bibfnamefont {R.}~\bibnamefont {Scelle}},
  \bibinfo {author} {\bibfnamefont {A.}~\bibnamefont {Trautmann}}, \bibinfo
  {author} {\bibfnamefont {S.}~\bibnamefont {Knoop}}, \bibinfo {author}
  {\bibfnamefont {M.~K.}\ \bibnamefont {Oberthaler}}, \bibinfo {author}
  {\bibfnamefont {M.~M.}\ \bibnamefont {Haverhals}}, \bibinfo {author}
  {\bibfnamefont {M.~R.}\ \bibnamefont {Goosen}}, \bibinfo {author}
  {\bibfnamefont {S.~J. J. M.~F.}\ \bibnamefont {Kokkelmans}}, \ and\ \bibinfo
  {author} {\bibfnamefont {E.}~\bibnamefont {Tiemann}},\ }\href@noop {}
  {\bibfield  {journal} {\bibinfo  {journal} {Phys. Rev. A}\ }\textbf {\bibinfo
  {volume} {85}},\ \bibinfo {pages} {042721} (\bibinfo {year}
  {2012})}\BibitemShut {NoStop}%
\bibitem [{\citenamefont {Tung}\ \emph {et~al.}(2013)\citenamefont {Tung},
  \citenamefont {Parker}, \citenamefont {Johansen}, \citenamefont {Chin},
  \citenamefont {Wang},\ and\ \citenamefont {Julienne}}]{Tung-PRA13}%
  \BibitemOpen
  \bibfield  {author} {\bibinfo {author} {\bibfnamefont {S.~K.}\ \bibnamefont
  {Tung}}, \bibinfo {author} {\bibfnamefont {C.}~\bibnamefont {Parker}},
  \bibinfo {author} {\bibfnamefont {J.}~\bibnamefont {Johansen}}, \bibinfo
  {author} {\bibfnamefont {C.}~\bibnamefont {Chin}}, \bibinfo {author}
  {\bibfnamefont {Y.}~\bibnamefont {Wang}}, \ and\ \bibinfo {author}
  {\bibfnamefont {P.~S.}\ \bibnamefont {Julienne}},\ }\href@noop {} {\bibfield
  {journal} {\bibinfo  {journal} {Phys. Rev. A}\ }\textbf {\bibinfo {volume}
  {87}},\ \bibinfo {pages} {010702(R)} (\bibinfo {year} {2013})}\BibitemShut
  {NoStop}%
\bibitem [{\citenamefont {Ferrier-Barbut}\ \emph {et~al.}(2014)\citenamefont
  {Ferrier-Barbut}, \citenamefont {Delehaye}, \citenamefont {Laurent},
  \citenamefont {Grier}, \citenamefont {Pierce}, \citenamefont {Rem},
  \citenamefont {Chevy},\ and\ \citenamefont {Salomon}}]{Ferrier-Barbut-S14}%
  \BibitemOpen
  \bibfield  {author} {\bibinfo {author} {\bibfnamefont {I.}~\bibnamefont
  {Ferrier-Barbut}}, \bibinfo {author} {\bibfnamefont {M.}~\bibnamefont
  {Delehaye}}, \bibinfo {author} {\bibfnamefont {S.}~\bibnamefont {Laurent}},
  \bibinfo {author} {\bibfnamefont {A.~T.}\ \bibnamefont {Grier}}, \bibinfo
  {author} {\bibfnamefont {M.}~\bibnamefont {Pierce}}, \bibinfo {author}
  {\bibfnamefont {B.~S.}\ \bibnamefont {Rem}}, \bibinfo {author} {\bibfnamefont
  {F.}~\bibnamefont {Chevy}}, \ and\ \bibinfo {author} {\bibfnamefont
  {C.}~\bibnamefont {Salomon}},\ }\href@noop {} {\bibfield  {journal} {\bibinfo
   {journal} {Science}\ }\textbf {\bibinfo {volume} {345}},\ \bibinfo {pages}
  {1035} (\bibinfo {year} {2014})}\BibitemShut {NoStop}%
\bibitem [{\citenamefont {Delehaye}\ \emph {et~al.}(2015)\citenamefont
  {Delehaye}, \citenamefont {Laurent}, \citenamefont {Ferrier-Barbut},
  \citenamefont {Jin}, \citenamefont {Chevy},\ and\ \citenamefont
  {Salomon}}]{Delehaye-PRL15}%
  \BibitemOpen
  \bibfield  {author} {\bibinfo {author} {\bibfnamefont {M.}~\bibnamefont
  {Delehaye}}, \bibinfo {author} {\bibfnamefont {S.}~\bibnamefont {Laurent}},
  \bibinfo {author} {\bibfnamefont {I.}~\bibnamefont {Ferrier-Barbut}},
  \bibinfo {author} {\bibfnamefont {S.}~\bibnamefont {Jin}}, \bibinfo {author}
  {\bibfnamefont {F.}~\bibnamefont {Chevy}}, \ and\ \bibinfo {author}
  {\bibfnamefont {C.}~\bibnamefont {Salomon}},\ }\href@noop {} {\bibfield
  {journal} {\bibinfo  {journal} {Phys. Rev. Lett.}\ }\textbf {\bibinfo
  {volume} {115}},\ \bibinfo {pages} {265303} (\bibinfo {year}
  {2015})}\BibitemShut {NoStop}%
\bibitem [{\citenamefont {Vaidya}\ \emph {et~al.}(2015)\citenamefont {Vaidya},
  \citenamefont {Tiamsuphat}, \citenamefont {Rolston},\ and\ \citenamefont
  {Porto}}]{Vaidya-PRA15}%
  \BibitemOpen
  \bibfield  {author} {\bibinfo {author} {\bibfnamefont {V.~D.}\ \bibnamefont
  {Vaidya}}, \bibinfo {author} {\bibfnamefont {J.}~\bibnamefont {Tiamsuphat}},
  \bibinfo {author} {\bibfnamefont {S.~L.}\ \bibnamefont {Rolston}}, \ and\
  \bibinfo {author} {\bibfnamefont {J.~V.}\ \bibnamefont {Porto}},\ }\href@noop
  {} {\bibfield  {journal} {\bibinfo  {journal} {Phys. Rev. A}\ }\textbf
  {\bibinfo {volume} {92}},\ \bibinfo {pages} {043604} (\bibinfo {year}
  {2015})}\BibitemShut {NoStop}%
\bibitem [{\citenamefont {Yao}\ \emph {et~al.}(2016)\citenamefont {Yao},
  \citenamefont {Chen}, \citenamefont {Wu}, \citenamefont {Liu}, \citenamefont
  {Wang}, \citenamefont {Jiang}, \citenamefont {Deng}, \citenamefont {Chen},\
  and\ \citenamefont {Pan}}]{XCYao-PRL16}%
  \BibitemOpen
  \bibfield  {author} {\bibinfo {author} {\bibfnamefont {X.-C.}\ \bibnamefont
  {Yao}}, \bibinfo {author} {\bibfnamefont {H.-Z.}\ \bibnamefont {Chen}},
  \bibinfo {author} {\bibfnamefont {Y.-P.}\ \bibnamefont {Wu}}, \bibinfo
  {author} {\bibfnamefont {X.-P.}\ \bibnamefont {Liu}}, \bibinfo {author}
  {\bibfnamefont {X.-Q.}\ \bibnamefont {Wang}}, \bibinfo {author}
  {\bibfnamefont {X.}~\bibnamefont {Jiang}}, \bibinfo {author} {\bibfnamefont
  {Y.}~\bibnamefont {Deng}}, \bibinfo {author} {\bibfnamefont {Y.-A.}\
  \bibnamefont {Chen}}, \ and\ \bibinfo {author} {\bibfnamefont {J.-W.}\
  \bibnamefont {Pan}},\ }\href@noop {} {\bibfield  {journal} {\bibinfo
  {journal} {Phys. Rev. Lett.}\ }\textbf {\bibinfo {volume} {117}},\ \bibinfo
  {pages} {145301} (\bibinfo {year} {2016})}\BibitemShut {NoStop}%
\bibitem [{\citenamefont {Onofrio}(2016)}]{Onofrio-PUsp16}%
  \BibitemOpen
  \bibfield  {author} {\bibinfo {author} {\bibfnamefont {R.}~\bibnamefont
  {Onofrio}},\ }\href@noop {} {\bibfield  {journal} {\bibinfo  {journal} {Phys.
  Usp.}\ }\textbf {\bibinfo {volume} {59}},\ \bibinfo {pages} {1129} (\bibinfo
  {year} {2016})}\BibitemShut {NoStop}%
\bibitem [{\citenamefont {Wu}\ \emph {et~al.}(2017)\citenamefont {Wu},
  \citenamefont {Yao}, \citenamefont {Chen}, \citenamefont {Liu}, \citenamefont
  {Wang}, \citenamefont {Chen},\ and\ \citenamefont {Pan}}]{YPWu-JPB17}%
  \BibitemOpen
  \bibfield  {author} {\bibinfo {author} {\bibfnamefont {Y.-P.}\ \bibnamefont
  {Wu}}, \bibinfo {author} {\bibfnamefont {X.-C.}\ \bibnamefont {Yao}},
  \bibinfo {author} {\bibfnamefont {H.-Z.}\ \bibnamefont {Chen}}, \bibinfo
  {author} {\bibfnamefont {X.-P.}\ \bibnamefont {Liu}}, \bibinfo {author}
  {\bibfnamefont {X.-Q.}\ \bibnamefont {Wang}}, \bibinfo {author}
  {\bibfnamefont {Y.-A.}\ \bibnamefont {Chen}}, \ and\ \bibinfo {author}
  {\bibfnamefont {J.-W.}\ \bibnamefont {Pan}},\ }\href@noop {} {\bibfield
  {journal} {\bibinfo  {journal} {J. Phys. B: At. Mol. Opt. Phys.}\ }\textbf
  {\bibinfo {volume} {50}},\ \bibinfo {pages} {094001} (\bibinfo {year}
  {2017})}\BibitemShut {NoStop}%
\bibitem [{\citenamefont {Roy}\ \emph {et~al.}(2017)\citenamefont {Roy},
  \citenamefont {Green}, \citenamefont {Bowler},\ and\ \citenamefont
  {Gupta}}]{Roy-PRL17}%
  \BibitemOpen
  \bibfield  {author} {\bibinfo {author} {\bibfnamefont {R.}~\bibnamefont
  {Roy}}, \bibinfo {author} {\bibfnamefont {A.}~\bibnamefont {Green}}, \bibinfo
  {author} {\bibfnamefont {R.}~\bibnamefont {Bowler}}, \ and\ \bibinfo {author}
  {\bibfnamefont {S.}~\bibnamefont {Gupta}},\ }\href@noop {} {\bibfield
  {journal} {\bibinfo  {journal} {Phys. Rev. Lett.}\ }\textbf {\bibinfo
  {volume} {118}},\ \bibinfo {pages} {055301} (\bibinfo {year}
  {2017})}\BibitemShut {NoStop}%
\bibitem [{\citenamefont {Sch{\"a}fer}\ \emph {et~al.}(2018)\citenamefont
  {Sch{\"a}fer}, \citenamefont {Mizukami}, \citenamefont {Yu}, \citenamefont
  {Koibuchi}, \citenamefont {Bouscal},\ and\ \citenamefont
  {Takahashi}}]{Schafer-PRA18}%
  \BibitemOpen
  \bibfield  {author} {\bibinfo {author} {\bibfnamefont {F.}~\bibnamefont
  {Sch{\"a}fer}}, \bibinfo {author} {\bibfnamefont {N.}~\bibnamefont
  {Mizukami}}, \bibinfo {author} {\bibfnamefont {P.}~\bibnamefont {Yu}},
  \bibinfo {author} {\bibfnamefont {S.}~\bibnamefont {Koibuchi}}, \bibinfo
  {author} {\bibfnamefont {A.}~\bibnamefont {Bouscal}}, \ and\ \bibinfo
  {author} {\bibfnamefont {Y.}~\bibnamefont {Takahashi}},\ }\href@noop {}
  {\bibfield  {journal} {\bibinfo  {journal} {Phys. Rev. A}\ }\textbf {\bibinfo
  {volume} {98}},\ \bibinfo {pages} {051602(R)} (\bibinfo {year}
  {2018})}\BibitemShut {NoStop}%
\bibitem [{\citenamefont {Lous}\ \emph {et~al.}(2018)\citenamefont {Lous},
  \citenamefont {Fritsche}, \citenamefont {Jag}, \citenamefont {Lehmann},
  \citenamefont {Kirilov}, \citenamefont {Huang},\ and\ \citenamefont
  {Grimm}}]{Lous-PRL18}%
  \BibitemOpen
  \bibfield  {author} {\bibinfo {author} {\bibfnamefont {R.~S.}\ \bibnamefont
  {Lous}}, \bibinfo {author} {\bibfnamefont {I.}~\bibnamefont {Fritsche}},
  \bibinfo {author} {\bibfnamefont {M.}~\bibnamefont {Jag}}, \bibinfo {author}
  {\bibfnamefont {F.}~\bibnamefont {Lehmann}}, \bibinfo {author} {\bibfnamefont
  {E.}~\bibnamefont {Kirilov}}, \bibinfo {author} {\bibfnamefont
  {B.}~\bibnamefont {Huang}}, \ and\ \bibinfo {author} {\bibfnamefont
  {R.}~\bibnamefont {Grimm}},\ }\href@noop {} {\bibfield  {journal} {\bibinfo
  {journal} {Phys. Rev. Lett.}\ }\textbf {\bibinfo {volume} {120}},\ \bibinfo
  {pages} {243403} (\bibinfo {year} {2018})}\BibitemShut {NoStop}%
\bibitem [{\citenamefont {Trautmann}\ \emph {et~al.}(2018)\citenamefont
  {Trautmann}, \citenamefont {Ilzh{\"o}fer}, \citenamefont {Durastante},
  \citenamefont {Politi}, \citenamefont {Sohmen}, \citenamefont {Mark},\ and\
  \citenamefont {Ferlaino}}]{Trautmann-PRL18}%
  \BibitemOpen
  \bibfield  {author} {\bibinfo {author} {\bibfnamefont {A.}~\bibnamefont
  {Trautmann}}, \bibinfo {author} {\bibfnamefont {P.}~\bibnamefont
  {Ilzh{\"o}fer}}, \bibinfo {author} {\bibfnamefont {G.}~\bibnamefont
  {Durastante}}, \bibinfo {author} {\bibfnamefont {C.}~\bibnamefont {Politi}},
  \bibinfo {author} {\bibfnamefont {M.}~\bibnamefont {Sohmen}}, \bibinfo
  {author} {\bibfnamefont {M.~J.}\ \bibnamefont {Mark}}, \ and\ \bibinfo
  {author} {\bibfnamefont {F.}~\bibnamefont {Ferlaino}},\ }\href@noop {}
  {\bibfield  {journal} {\bibinfo  {journal} {Phys. Rev. Lett.}\ }\textbf
  {\bibinfo {volume} {121}},\ \bibinfo {pages} {213601} (\bibinfo {year}
  {2018})}\BibitemShut {NoStop}%
\bibitem [{\citenamefont {Albus}\ \emph {et~al.}(2003)\citenamefont {Albus},
  \citenamefont {Illuminate},\ and\ \citenamefont {Eisert}}]{Albus-PRA03}%
  \BibitemOpen
  \bibfield  {author} {\bibinfo {author} {\bibfnamefont {A.}~\bibnamefont
  {Albus}}, \bibinfo {author} {\bibfnamefont {F.}~\bibnamefont {Illuminate}}, \
  and\ \bibinfo {author} {\bibfnamefont {J.}~\bibnamefont {Eisert}},\
  }\href@noop {} {\bibfield  {journal} {\bibinfo  {journal} {Phys. Rev. A}\
  }\textbf {\bibinfo {volume} {68}},\ \bibinfo {pages} {023606.} (\bibinfo
  {year} {2003})}\BibitemShut {NoStop}%
\bibitem [{\citenamefont {Cazalilla}\ and\ \citenamefont
  {Ho}(2003)}]{Cazalilla-PRL03}%
  \BibitemOpen
  \bibfield  {author} {\bibinfo {author} {\bibfnamefont {M.}~\bibnamefont
  {Cazalilla}}\ and\ \bibinfo {author} {\bibfnamefont {A.}~\bibnamefont {Ho}},\
  }\href@noop {} {\bibfield  {journal} {\bibinfo  {journal} {Phys. Rev. Lett.}\
  }\textbf {\bibinfo {volume} {91}},\ \bibinfo {pages} {150403} (\bibinfo
  {year} {2003})}\BibitemShut {NoStop}%
\bibitem [{\citenamefont {Lewenstein}\ \emph {et~al.}(2004)\citenamefont
  {Lewenstein}, \citenamefont {Santos}, \citenamefont {Baranov},\ and\
  \citenamefont {Fehrmann}}]{Lewenstein-PRL04}%
  \BibitemOpen
  \bibfield  {author} {\bibinfo {author} {\bibfnamefont {M.}~\bibnamefont
  {Lewenstein}}, \bibinfo {author} {\bibfnamefont {L.}~\bibnamefont {Santos}},
  \bibinfo {author} {\bibfnamefont {M.~A.}\ \bibnamefont {Baranov}}, \ and\
  \bibinfo {author} {\bibfnamefont {H.}~\bibnamefont {Fehrmann}},\ }\href@noop
  {} {\bibfield  {journal} {\bibinfo  {journal} {Phys. Rev. Lett.}\ }\textbf
  {\bibinfo {volume} {92}},\ \bibinfo {pages} {050401} (\bibinfo {year}
  {2004})}\BibitemShut {NoStop}%
\bibitem [{\citenamefont {Mathey}\ \emph {et~al.}(2004)\citenamefont {Mathey},
  \citenamefont {Wang}, \citenamefont {Hofstetter}, \citenamefont {Lukin},\
  and\ \citenamefont {Demler}}]{Mathey-PRL04}%
  \BibitemOpen
  \bibfield  {author} {\bibinfo {author} {\bibfnamefont {L.}~\bibnamefont
  {Mathey}}, \bibinfo {author} {\bibfnamefont {D.-W.}\ \bibnamefont {Wang}},
  \bibinfo {author} {\bibfnamefont {W.}~\bibnamefont {Hofstetter}}, \bibinfo
  {author} {\bibfnamefont {M.~D.}\ \bibnamefont {Lukin}}, \ and\ \bibinfo
  {author} {\bibfnamefont {E.}~\bibnamefont {Demler}},\ }\href@noop {}
  {\bibfield  {journal} {\bibinfo  {journal} {Phys. Rev. Lett.}\ }\textbf
  {\bibinfo {volume} {93}},\ \bibinfo {pages} {120404} (\bibinfo {year}
  {2004})}\BibitemShut {NoStop}%
\bibitem [{\citenamefont {Roth}\ and\ \citenamefont
  {Burnett}(2004)}]{Roth-PRA04}%
  \BibitemOpen
  \bibfield  {author} {\bibinfo {author} {\bibfnamefont {R.}~\bibnamefont
  {Roth}}\ and\ \bibinfo {author} {\bibfnamefont {K.}~\bibnamefont {Burnett}},\
  }\href@noop {} {\bibfield  {journal} {\bibinfo  {journal} {Phys. Rev. A}\
  }\textbf {\bibinfo {volume} {69}},\ \bibinfo {pages} {021601} (\bibinfo
  {year} {2004})}\BibitemShut {NoStop}%
\bibitem [{\citenamefont {Frahm}\ and\ \citenamefont
  {Palacios}(2005)}]{Frahm-PRA05}%
  \BibitemOpen
  \bibfield  {author} {\bibinfo {author} {\bibfnamefont {H.}~\bibnamefont
  {Frahm}}\ and\ \bibinfo {author} {\bibfnamefont {G.}~\bibnamefont
  {Palacios}},\ }\href@noop {} {\bibfield  {journal} {\bibinfo  {journal}
  {Phys. Rev. A}\ }\textbf {\bibinfo {volume} {72}},\ \bibinfo {pages} {061604}
  (\bibinfo {year} {2005})}\BibitemShut {NoStop}%
\bibitem [{\citenamefont {Batchelor}\ \emph {et~al.}(2005)\citenamefont
  {Batchelor}, \citenamefont {Bortz}, \citenamefont {Guan},\ and\ \citenamefont
  {Oelkers}}]{Batchelor-PRA05}%
  \BibitemOpen
  \bibfield  {author} {\bibinfo {author} {\bibfnamefont {M.~T.}\ \bibnamefont
  {Batchelor}}, \bibinfo {author} {\bibfnamefont {M.}~\bibnamefont {Bortz}},
  \bibinfo {author} {\bibfnamefont {X.~W.}\ \bibnamefont {Guan}}, \ and\
  \bibinfo {author} {\bibfnamefont {N.}~\bibnamefont {Oelkers}},\ }\href@noop
  {} {\bibfield  {journal} {\bibinfo  {journal} {Phys. Rev. A}\ }\textbf
  {\bibinfo {volume} {72}},\ \bibinfo {pages} {061603} (\bibinfo {year}
  {2005})}\BibitemShut {NoStop}%
\bibitem [{\citenamefont {Takeuchi}\ and\ \citenamefont
  {Mori}(2005)}]{Takeuchi-PRA05}%
  \BibitemOpen
  \bibfield  {author} {\bibinfo {author} {\bibfnamefont {Y.}~\bibnamefont
  {Takeuchi}}\ and\ \bibinfo {author} {\bibfnamefont {H.}~\bibnamefont
  {Mori}},\ }\href@noop {} {\bibfield  {journal} {\bibinfo  {journal} {Phys.
  Rev. A}\ }\textbf {\bibinfo {volume} {72}},\ \bibinfo {pages} {063617}
  (\bibinfo {year} {2005})}\BibitemShut {NoStop}%
\bibitem [{\citenamefont {Pollet}\ \emph {et~al.}(2006)\citenamefont {Pollet},
  \citenamefont {Troyer}, \citenamefont {Houcke},\ and\ \citenamefont
  {Rombouts}}]{Pollet-PRL06}%
  \BibitemOpen
  \bibfield  {author} {\bibinfo {author} {\bibfnamefont {L.}~\bibnamefont
  {Pollet}}, \bibinfo {author} {\bibfnamefont {M.}~\bibnamefont {Troyer}},
  \bibinfo {author} {\bibfnamefont {K.~V.}\ \bibnamefont {Houcke}}, \ and\
  \bibinfo {author} {\bibfnamefont {S.}~\bibnamefont {Rombouts}},\ }\href@noop
  {} {\bibfield  {journal} {\bibinfo  {journal} {Phys. Rev. Lett.}\ }\textbf
  {\bibinfo {volume} {96}},\ \bibinfo {pages} {190402} (\bibinfo {year}
  {2006})}\BibitemShut {NoStop}%
\bibitem [{\citenamefont {Mathey}\ and\ \citenamefont
  {Wang}(2007)}]{Mathey-PRA07}%
  \BibitemOpen
  \bibfield  {author} {\bibinfo {author} {\bibfnamefont {L.}~\bibnamefont
  {Mathey}}\ and\ \bibinfo {author} {\bibfnamefont {D.-W.}\ \bibnamefont
  {Wang}},\ }\href@noop {} {\bibfield  {journal} {\bibinfo  {journal} {Phys.
  Rev. A}\ }\textbf {\bibinfo {volume} {75}},\ \bibinfo {pages} {013612}
  (\bibinfo {year} {2007})}\BibitemShut {NoStop}%
\bibitem [{\citenamefont {Sengupta}\ \emph {et~al.}(2007)\citenamefont
  {Sengupta}, \citenamefont {Dupuis},\ and\ \citenamefont
  {Majumdar}}]{Sengupta-PRA07}%
  \BibitemOpen
  \bibfield  {author} {\bibinfo {author} {\bibfnamefont {K.}~\bibnamefont
  {Sengupta}}, \bibinfo {author} {\bibfnamefont {N.}~\bibnamefont {Dupuis}}, \
  and\ \bibinfo {author} {\bibfnamefont {P.}~\bibnamefont {Majumdar}},\
  }\href@noop {} {\bibfield  {journal} {\bibinfo  {journal} {Phys. Rev. A}\
  }\textbf {\bibinfo {volume} {75}},\ \bibinfo {pages} {063625.} (\bibinfo
  {year} {2007})}\BibitemShut {NoStop}%
\bibitem [{\citenamefont {Mering}\ and\ \citenamefont
  {Fleischhauer}(2008)}]{Mering-PRA08}%
  \BibitemOpen
  \bibfield  {author} {\bibinfo {author} {\bibfnamefont {A.}~\bibnamefont
  {Mering}}\ and\ \bibinfo {author} {\bibfnamefont {M.}~\bibnamefont
  {Fleischhauer}},\ }\href@noop {} {\bibfield  {journal} {\bibinfo  {journal}
  {Phys. Rev. A}\ }\textbf {\bibinfo {volume} {77}},\ \bibinfo {pages} {023601}
  (\bibinfo {year} {2008})}\BibitemShut {NoStop}%
\bibitem [{\citenamefont {Suzuki}\ \emph {et~al.}(2008)\citenamefont {Suzuki},
  \citenamefont {Miyakawa},\ and\ \citenamefont {Suzuki}}]{Suzuki-PRA08}%
  \BibitemOpen
  \bibfield  {author} {\bibinfo {author} {\bibfnamefont {K.}~\bibnamefont
  {Suzuki}}, \bibinfo {author} {\bibfnamefont {T.}~\bibnamefont {Miyakawa}}, \
  and\ \bibinfo {author} {\bibfnamefont {T.}~\bibnamefont {Suzuki}},\
  }\href@noop {} {\bibfield  {journal} {\bibinfo  {journal} {Phys. Rev. A}\
  }\textbf {\bibinfo {volume} {77}},\ \bibinfo {pages} {043629} (\bibinfo
  {year} {2008})}\BibitemShut {NoStop}%
\bibitem [{\citenamefont {L{\"u}hmann}\ \emph {et~al.}(2008)\citenamefont
  {L{\"u}hmann}, \citenamefont {Bongs}, \citenamefont {Sengstock},\ and\
  \citenamefont {Pfannkuche}}]{Luhmann-PRL08}%
  \BibitemOpen
  \bibfield  {author} {\bibinfo {author} {\bibfnamefont {D.-S.}\ \bibnamefont
  {L{\"u}hmann}}, \bibinfo {author} {\bibfnamefont {K.}~\bibnamefont {Bongs}},
  \bibinfo {author} {\bibfnamefont {K.}~\bibnamefont {Sengstock}}, \ and\
  \bibinfo {author} {\bibfnamefont {D.}~\bibnamefont {Pfannkuche}},\
  }\href@noop {} {\bibfield  {journal} {\bibinfo  {journal} {Phys. Rev. Lett.}\
  }\textbf {\bibinfo {volume} {101}},\ \bibinfo {pages} {050402} (\bibinfo
  {year} {2008})}\BibitemShut {NoStop}%
\bibitem [{\citenamefont {Rizzi}\ and\ \citenamefont
  {Imambekov}(2008)}]{Rizzi-PRA08}%
  \BibitemOpen
  \bibfield  {author} {\bibinfo {author} {\bibfnamefont {M.}~\bibnamefont
  {Rizzi}}\ and\ \bibinfo {author} {\bibfnamefont {A.}~\bibnamefont
  {Imambekov}},\ }\href@noop {} {\bibfield  {journal} {\bibinfo  {journal}
  {Phys. Rev. A}\ }\textbf {\bibinfo {volume} {77}},\ \bibinfo {pages} {023621}
  (\bibinfo {year} {2008})}\BibitemShut {NoStop}%
\bibitem [{\citenamefont {Orth}\ \emph {et~al.}(2009)\citenamefont {Orth},
  \citenamefont {Bergman},\ and\ \citenamefont {Hur}}]{Orth-PRA09}%
  \BibitemOpen
  \bibfield  {author} {\bibinfo {author} {\bibfnamefont {P.~P.}\ \bibnamefont
  {Orth}}, \bibinfo {author} {\bibfnamefont {D.~L.}\ \bibnamefont {Bergman}}, \
  and\ \bibinfo {author} {\bibfnamefont {K.~L.}\ \bibnamefont {Hur}},\
  }\href@noop {} {\bibfield  {journal} {\bibinfo  {journal} {Phys. Rev. A}\
  }\textbf {\bibinfo {volume} {80}},\ \bibinfo {pages} {023624} (\bibinfo
  {year} {2009})}\BibitemShut {NoStop}%
\bibitem [{\citenamefont {Yin}\ \emph {et~al.}(2009)\citenamefont {Yin},
  \citenamefont {Chen},\ and\ \citenamefont {Zhang}}]{XYin-PRA09}%
  \BibitemOpen
  \bibfield  {author} {\bibinfo {author} {\bibfnamefont {X.}~\bibnamefont
  {Yin}}, \bibinfo {author} {\bibfnamefont {S.}~\bibnamefont {Chen}}, \ and\
  \bibinfo {author} {\bibfnamefont {Y.}~\bibnamefont {Zhang}},\ }\href@noop {}
  {\bibfield  {journal} {\bibinfo  {journal} {Phys. Rev. A}\ }\textbf {\bibinfo
  {volume} {79}},\ \bibinfo {pages} {053604} (\bibinfo {year}
  {2009})}\BibitemShut {NoStop}%
\bibitem [{\citenamefont {Sinha}\ and\ \citenamefont
  {Sengupta}(2009)}]{Sinha-PRB09}%
  \BibitemOpen
  \bibfield  {author} {\bibinfo {author} {\bibfnamefont {S.}~\bibnamefont
  {Sinha}}\ and\ \bibinfo {author} {\bibfnamefont {K.}~\bibnamefont
  {Sengupta}},\ }\href@noop {} {\bibfield  {journal} {\bibinfo  {journal}
  {Phys. Rev. B}\ }\textbf {\bibinfo {volume} {79}},\ \bibinfo {pages} {115124}
  (\bibinfo {year} {2009})}\BibitemShut {NoStop}%
\bibitem [{\citenamefont {Orignac}\ \emph {et~al.}(2010)\citenamefont
  {Orignac}, \citenamefont {Tsuchiizu},\ and\ \citenamefont
  {Suzumura}}]{Orignac-PRA10}%
  \BibitemOpen
  \bibfield  {author} {\bibinfo {author} {\bibfnamefont {E.}~\bibnamefont
  {Orignac}}, \bibinfo {author} {\bibfnamefont {M.}~\bibnamefont {Tsuchiizu}},
  \ and\ \bibinfo {author} {\bibfnamefont {Y.}~\bibnamefont {Suzumura}},\
  }\href@noop {} {\bibfield  {journal} {\bibinfo  {journal} {Phys. Rev. A}\
  }\textbf {\bibinfo {volume} {81}},\ \bibinfo {pages} {053626} (\bibinfo
  {year} {2010})}\BibitemShut {NoStop}%
\bibitem [{\citenamefont {Polak}\ and\ \citenamefont
  {Kope{\'c}}(2010)}]{Polak-PRA10}%
  \BibitemOpen
  \bibfield  {author} {\bibinfo {author} {\bibfnamefont {T.~P.}\ \bibnamefont
  {Polak}}\ and\ \bibinfo {author} {\bibfnamefont {T.~K.}\ \bibnamefont
  {Kope{\'c}}},\ }\href@noop {} {\bibfield  {journal} {\bibinfo  {journal}
  {Phys. Rev. A}\ }\textbf {\bibinfo {volume} {81}},\ \bibinfo {pages} {043612}
  (\bibinfo {year} {2010})}\BibitemShut {NoStop}%
\bibitem [{\citenamefont {Mering}\ and\ \citenamefont
  {Fleischhauer}(2010)}]{Mering-PRA10}%
  \BibitemOpen
  \bibfield  {author} {\bibinfo {author} {\bibfnamefont {A.}~\bibnamefont
  {Mering}}\ and\ \bibinfo {author} {\bibfnamefont {M.}~\bibnamefont
  {Fleischhauer}},\ }\href@noop {} {\bibfield  {journal} {\bibinfo  {journal}
  {Phys. Rev. A}\ }\textbf {\bibinfo {volume} {81}},\ \bibinfo {pages}
  {011603(R)} (\bibinfo {year} {2010})}\BibitemShut {NoStop}%
\bibitem [{\citenamefont {Anders}\ \emph {et~al.}(2012)\citenamefont {Anders},
  \citenamefont {Werner}, \citenamefont {Troyer}, \citenamefont {Sigrist},\
  and\ \citenamefont {Pollet}}]{Anders-PRL12}%
  \BibitemOpen
  \bibfield  {author} {\bibinfo {author} {\bibfnamefont {P.}~\bibnamefont
  {Anders}}, \bibinfo {author} {\bibfnamefont {P.}~\bibnamefont {Werner}},
  \bibinfo {author} {\bibfnamefont {M.}~\bibnamefont {Troyer}}, \bibinfo
  {author} {\bibfnamefont {M.}~\bibnamefont {Sigrist}}, \ and\ \bibinfo
  {author} {\bibfnamefont {L.}~\bibnamefont {Pollet}},\ }\href@noop {}
  {\bibfield  {journal} {\bibinfo  {journal} {Phys. Rev. Lett.}\ }\textbf
  {\bibinfo {volume} {109}},\ \bibinfo {pages} {206401} (\bibinfo {year}
  {2012})}\BibitemShut {NoStop}%
\bibitem [{\citenamefont {Masaki}\ and\ \citenamefont
  {Mori}(2013)}]{Masaki-JPSJ13}%
  \BibitemOpen
  \bibfield  {author} {\bibinfo {author} {\bibfnamefont {A.}~\bibnamefont
  {Masaki}}\ and\ \bibinfo {author} {\bibfnamefont {H.}~\bibnamefont {Mori}},\
  }\href@noop {} {\bibfield  {journal} {\bibinfo  {journal} {J. Phys. Soc.
  Jpn.}\ }\textbf {\bibinfo {volume} {82}},\ \bibinfo {pages} {074002}
  (\bibinfo {year} {2013})}\BibitemShut {NoStop}%
\bibitem [{\citenamefont {Bukov}\ and\ \citenamefont
  {Pollet}(2014)}]{Bukov-PRB14}%
  \BibitemOpen
  \bibfield  {author} {\bibinfo {author} {\bibfnamefont {M.}~\bibnamefont
  {Bukov}}\ and\ \bibinfo {author} {\bibfnamefont {L.}~\bibnamefont {Pollet}},\
  }\href@noop {} {\bibfield  {journal} {\bibinfo  {journal} {Phys. Rev. B}\
  }\textbf {\bibinfo {volume} {89}},\ \bibinfo {pages} {094502} (\bibinfo
  {year} {2014})}\BibitemShut {NoStop}%
\bibitem [{\citenamefont {Ozawa}\ \emph {et~al.}(2014)\citenamefont {Ozawa},
  \citenamefont {Recati}, \citenamefont {Delehaye}, \citenamefont {Chevy},\
  and\ \citenamefont {Stringari}}]{TOzawa-PRA14}%
  \BibitemOpen
  \bibfield  {author} {\bibinfo {author} {\bibfnamefont {T.}~\bibnamefont
  {Ozawa}}, \bibinfo {author} {\bibfnamefont {A.}~\bibnamefont {Recati}},
  \bibinfo {author} {\bibfnamefont {M.}~\bibnamefont {Delehaye}}, \bibinfo
  {author} {\bibfnamefont {F.}~\bibnamefont {Chevy}}, \ and\ \bibinfo {author}
  {\bibfnamefont {S.}~\bibnamefont {Stringari}},\ }\href@noop {} {\bibfield
  {journal} {\bibinfo  {journal} {Phys. Rev. A}\ }\textbf {\bibinfo {volume}
  {90}},\ \bibinfo {pages} {043608} (\bibinfo {year} {2014})}\BibitemShut
  {NoStop}%
\bibitem [{\citenamefont {Bilitewski}\ and\ \citenamefont
  {Pollet}(2015)}]{Bilitewski-PRB15}%
  \BibitemOpen
  \bibfield  {author} {\bibinfo {author} {\bibfnamefont {T.}~\bibnamefont
  {Bilitewski}}\ and\ \bibinfo {author} {\bibfnamefont {L.}~\bibnamefont
  {Pollet}},\ }\href@noop {} {\bibfield  {journal} {\bibinfo  {journal} {Phys.
  Rev. B}\ }\textbf {\bibinfo {volume} {92}},\ \bibinfo {pages} {184505}
  (\bibinfo {year} {2015})}\BibitemShut {NoStop}%
\bibitem [{\citenamefont {Zujev}\ \emph {et~al.}(2008)\citenamefont {Zujev},
  \citenamefont {Baldwin}, \citenamefont {Scalettar}, \citenamefont {Rousseau},
  \citenamefont {Denteneer},\ and\ \citenamefont {Rigol}}]{Zujev-PRA08}%
  \BibitemOpen
  \bibfield  {author} {\bibinfo {author} {\bibfnamefont {A.}~\bibnamefont
  {Zujev}}, \bibinfo {author} {\bibfnamefont {A.}~\bibnamefont {Baldwin}},
  \bibinfo {author} {\bibfnamefont {R.~T.}\ \bibnamefont {Scalettar}}, \bibinfo
  {author} {\bibfnamefont {V.~G.}\ \bibnamefont {Rousseau}}, \bibinfo {author}
  {\bibfnamefont {P.~J.~H.}\ \bibnamefont {Denteneer}}, \ and\ \bibinfo
  {author} {\bibfnamefont {M.}~\bibnamefont {Rigol}},\ }\href@noop {}
  {\bibfield  {journal} {\bibinfo  {journal} {Phys. Rev. A}\ }\textbf {\bibinfo
  {volume} {78}},\ \bibinfo {pages} {033619} (\bibinfo {year}
  {2008})}\BibitemShut {NoStop}%
\bibitem [{\citenamefont {Avella}\ \emph {et~al.}(2019)\citenamefont {Avella},
  \citenamefont {Mendoza-Arenas}, \citenamefont {Franco},\ and\ \citenamefont
  {Silva-Valencia}}]{Avella-PRA19}%
  \BibitemOpen
  \bibfield  {author} {\bibinfo {author} {\bibfnamefont {R.}~\bibnamefont
  {Avella}}, \bibinfo {author} {\bibfnamefont {J.~J.}\ \bibnamefont
  {Mendoza-Arenas}}, \bibinfo {author} {\bibfnamefont {R.}~\bibnamefont
  {Franco}}, \ and\ \bibinfo {author} {\bibfnamefont {J.}~\bibnamefont
  {Silva-Valencia}},\ }\href@noop {} {\bibfield  {journal} {\bibinfo  {journal}
  {Phys. Rev. A}\ }\textbf {\bibinfo {volume} {100}},\ \bibinfo {pages}
  {063620} (\bibinfo {year} {2019})}\BibitemShut {NoStop}%
\bibitem [{\citenamefont {Avella}\ \emph {et~al.}(2020)\citenamefont {Avella},
  \citenamefont {Mendoza-Arenas}, \citenamefont {Franco},\ and\ \citenamefont
  {Silva-Valencia}}]{Avella-PRA20}%
  \BibitemOpen
  \bibfield  {author} {\bibinfo {author} {\bibfnamefont {R.}~\bibnamefont
  {Avella}}, \bibinfo {author} {\bibfnamefont {J.~J.}\ \bibnamefont
  {Mendoza-Arenas}}, \bibinfo {author} {\bibfnamefont {R.}~\bibnamefont
  {Franco}}, \ and\ \bibinfo {author} {\bibfnamefont {J.}~\bibnamefont
  {Silva-Valencia}},\ }\href@noop {} {\bibfield  {journal} {\bibinfo  {journal}
  {Phys. Rev. A}\ }\textbf {\bibinfo {volume} {102}},\ \bibinfo {pages}
  {033341} (\bibinfo {year} {2020})}\BibitemShut {NoStop}%
\bibitem [{\citenamefont {Zwierlein}\ \emph {et~al.}(2006)\citenamefont
  {Zwierlein}, \citenamefont {Schirotzek}, \citenamefont {Schunck},\ and\
  \citenamefont {Ketterle}}]{Zwierlein-S06}%
  \BibitemOpen
  \bibfield  {author} {\bibinfo {author} {\bibfnamefont {M.~W.}\ \bibnamefont
  {Zwierlein}}, \bibinfo {author} {\bibfnamefont {A.}~\bibnamefont
  {Schirotzek}}, \bibinfo {author} {\bibfnamefont {C.~H.}\ \bibnamefont
  {Schunck}}, \ and\ \bibinfo {author} {\bibfnamefont {W.}~\bibnamefont
  {Ketterle}},\ }\href@noop {} {\bibfield  {journal} {\bibinfo  {journal}
  {Science}\ }\textbf {\bibinfo {volume} {311}},\ \bibinfo {pages} {492}
  (\bibinfo {year} {2006})}\BibitemShut {NoStop}%
\bibitem [{\citenamefont {Partridge}\ \emph {et~al.}(2006)\citenamefont
  {Partridge}, \citenamefont {Li}, \citenamefont {Kamar}, \citenamefont
  {Liao},\ and\ \citenamefont {Hulet}}]{Partridge-S06}%
  \BibitemOpen
  \bibfield  {author} {\bibinfo {author} {\bibfnamefont {G.~B.}\ \bibnamefont
  {Partridge}}, \bibinfo {author} {\bibfnamefont {W.~H.}\ \bibnamefont {Li}},
  \bibinfo {author} {\bibfnamefont {R.~I.}\ \bibnamefont {Kamar}}, \bibinfo
  {author} {\bibfnamefont {Y.~A.}\ \bibnamefont {Liao}}, \ and\ \bibinfo
  {author} {\bibfnamefont {R.~G.}\ \bibnamefont {Hulet}},\ }\href@noop {}
  {\bibfield  {journal} {\bibinfo  {journal} {Science}\ }\textbf {\bibinfo
  {volume} {311}},\ \bibinfo {pages} {503} (\bibinfo {year}
  {2006})}\BibitemShut {NoStop}%
\bibitem [{\citenamefont {a.~Liao}\ \emph {et~al.}(2010)\citenamefont
  {a.~Liao}, \citenamefont {Rittner}, \citenamefont {Paprotta}, \citenamefont
  {Li}, \citenamefont {Partridge}, \citenamefont {Hulet}, \citenamefont
  {Baur},\ and\ \citenamefont {Mueller}}]{Liao-Nat10}%
  \BibitemOpen
  \bibfield  {author} {\bibinfo {author} {\bibfnamefont {Y.}~\bibnamefont
  {a.~Liao}}, \bibinfo {author} {\bibfnamefont {A.~S.~C.}\ \bibnamefont
  {Rittner}}, \bibinfo {author} {\bibfnamefont {T.}~\bibnamefont {Paprotta}},
  \bibinfo {author} {\bibfnamefont {W.}~\bibnamefont {Li}}, \bibinfo {author}
  {\bibfnamefont {G.~B.}\ \bibnamefont {Partridge}}, \bibinfo {author}
  {\bibfnamefont {R.~G.}\ \bibnamefont {Hulet}}, \bibinfo {author}
  {\bibfnamefont {S.~K.}\ \bibnamefont {Baur}}, \ and\ \bibinfo {author}
  {\bibfnamefont {E.~J.}\ \bibnamefont {Mueller}},\ }\href@noop {} {\bibfield
  {journal} {\bibinfo  {journal} {Nature}\ }\textbf {\bibinfo {volume} {467}},\
  \bibinfo {pages} {567} (\bibinfo {year} {2010})}\BibitemShut {NoStop}%
\bibitem [{\citenamefont {Kinnunen}\ \emph {et~al.}(2018)\citenamefont
  {Kinnunen}, \citenamefont {Baarsma}, \citenamefont {Martikainen},\ and\
  \citenamefont {T{\"o}rm{\"a}}}]{Kinnunen-RPP18}%
  \BibitemOpen
  \bibfield  {author} {\bibinfo {author} {\bibfnamefont {J.~J.}\ \bibnamefont
  {Kinnunen}}, \bibinfo {author} {\bibfnamefont {J.~E.}\ \bibnamefont
  {Baarsma}}, \bibinfo {author} {\bibfnamefont {J.-P.}\ \bibnamefont
  {Martikainen}}, \ and\ \bibinfo {author} {\bibfnamefont {P.}~\bibnamefont
  {T{\"o}rm{\"a}}},\ }\href@noop {} {\bibfield  {journal} {\bibinfo  {journal}
  {Rep. Prog. Phys.}\ }\textbf {\bibinfo {volume} {81}},\ \bibinfo {pages}
  {046401} (\bibinfo {year} {2018})}\BibitemShut {NoStop}%
\bibitem [{\citenamefont {Dobrzyniecki}\ and\ \citenamefont
  {Sowi{\'n}ski}(2020)}]{Dobrzyniecki-AQT20}%
  \BibitemOpen
  \bibfield  {author} {\bibinfo {author} {\bibfnamefont {J.}~\bibnamefont
  {Dobrzyniecki}}\ and\ \bibinfo {author} {\bibfnamefont {T.}~\bibnamefont
  {Sowi{\'n}ski}},\ }\href@noop {} {\bibfield  {journal} {\bibinfo  {journal}
  {Adv. Quantum Technol.}\ }\textbf {\bibinfo {volume} {3}},\ \bibinfo {pages}
  {2000010} (\bibinfo {year} {2020})}\BibitemShut {NoStop}%
\bibitem [{\citenamefont {Fulde}\ and\ \citenamefont
  {Ferrell}(1964)}]{Fulde64}%
  \BibitemOpen
  \bibfield  {author} {\bibinfo {author} {\bibfnamefont {P.}~\bibnamefont
  {Fulde}}\ and\ \bibinfo {author} {\bibfnamefont {R.~A.}\ \bibnamefont
  {Ferrell}},\ }\href@noop {} {\bibfield  {journal} {\bibinfo  {journal} {Phys.
  Rev.}\ }\textbf {\bibinfo {volume} {135}},\ \bibinfo {pages} {A550} (\bibinfo
  {year} {1964})}\BibitemShut {NoStop}%
\bibitem [{\citenamefont {Larkin}\ and\ \citenamefont
  {Ovchinnikov}(1965)}]{Larkin65}%
  \BibitemOpen
  \bibfield  {author} {\bibinfo {author} {\bibfnamefont {A.~I.}\ \bibnamefont
  {Larkin}}\ and\ \bibinfo {author} {\bibfnamefont {Y.~N.}\ \bibnamefont
  {Ovchinnikov}},\ }\href@noop {} {\bibfield  {journal} {\bibinfo  {journal}
  {Sov. Phys.—JETP}\ }\textbf {\bibinfo {volume} {20}},\ \bibinfo {pages}
  {762} (\bibinfo {year} {1965})}\BibitemShut {NoStop}%
\bibitem [{\citenamefont {Singh}\ and\ \citenamefont
  {Orso}(2020)}]{Singh-PRR20}%
  \BibitemOpen
  \bibfield  {author} {\bibinfo {author} {\bibfnamefont {M.}~\bibnamefont
  {Singh}}\ and\ \bibinfo {author} {\bibfnamefont {G.}~\bibnamefont {Orso}},\
  }\href@noop {} {\bibfield  {journal} {\bibinfo  {journal} {Phys. Rev.
  Research}\ }\textbf {\bibinfo {volume} {2}},\ \bibinfo {pages} {023148}
  (\bibinfo {year} {2020})}\BibitemShut {NoStop}%
\bibitem [{\citenamefont {Sowi{\'n}ski}\ and\ \citenamefont
  {Garc{\'i}a-March}(2019)}]{Sowinski-RPP19}%
  \BibitemOpen
  \bibfield  {author} {\bibinfo {author} {\bibfnamefont {T.}~\bibnamefont
  {Sowi{\'n}ski}}\ and\ \bibinfo {author} {\bibfnamefont {M.~{\'A}.}\
  \bibnamefont {Garc{\'i}a-March}},\ }\href@noop {} {\bibfield  {journal}
  {\bibinfo  {journal} {Rep. Prog. Phys.}\ }\textbf {\bibinfo {volume} {82}},\
  \bibinfo {pages} {104401} (\bibinfo {year} {2019})}\BibitemShut {NoStop}%
\bibitem [{\citenamefont {Takasu}\ and\ \citenamefont
  {Takahashi}(2009)}]{YTakasu-JPSJ09}%
  \BibitemOpen
  \bibfield  {author} {\bibinfo {author} {\bibfnamefont {Y.}~\bibnamefont
  {Takasu}}\ and\ \bibinfo {author} {\bibfnamefont {Y.}~\bibnamefont
  {Takahashi}},\ }\href@noop {} {\bibfield  {journal} {\bibinfo  {journal} {J.
  Phys. Soc. Jpn.}\ }\textbf {\bibinfo {volume} {78}},\ \bibinfo {pages}
  {012001} (\bibinfo {year} {2009})}\BibitemShut {NoStop}%
\bibitem [{\citenamefont {Ikemachi}\ \emph {et~al.}(2017)\citenamefont
  {Ikemachi}, \citenamefont {Ito}, \citenamefont {Aratake}, \citenamefont
  {Chen}, \citenamefont {Koashi}, \citenamefont {Kuwata-Gonokami},\ and\
  \citenamefont {Horikoshi}}]{Ikemachi-JPB17}%
  \BibitemOpen
  \bibfield  {author} {\bibinfo {author} {\bibfnamefont {T.}~\bibnamefont
  {Ikemachi}}, \bibinfo {author} {\bibfnamefont {A.}~\bibnamefont {Ito}},
  \bibinfo {author} {\bibfnamefont {Y.}~\bibnamefont {Aratake}}, \bibinfo
  {author} {\bibfnamefont {Y.}~\bibnamefont {Chen}}, \bibinfo {author}
  {\bibfnamefont {M.}~\bibnamefont {Koashi}}, \bibinfo {author} {\bibfnamefont
  {M.}~\bibnamefont {Kuwata-Gonokami}}, \ and\ \bibinfo {author} {\bibfnamefont
  {M.}~\bibnamefont {Horikoshi}},\ }\href@noop {} {\bibfield  {journal}
  {\bibinfo  {journal} {J. Phys. B: At. Mol. Opt. Phys.}\ }\textbf {\bibinfo
  {volume} {50}},\ \bibinfo {pages} {01LT01} (\bibinfo {year}
  {2017})}\BibitemShut {NoStop}%
\bibitem [{\citenamefont {Cazalilla}\ \emph {et~al.}(2011)\citenamefont
  {Cazalilla}, \citenamefont {Citro}, \citenamefont {Giamarchi}, \citenamefont
  {Orignac},\ and\ \citenamefont {Rigol}}]{Cazalilla-RMP11}%
  \BibitemOpen
  \bibfield  {author} {\bibinfo {author} {\bibfnamefont {M.~A.}\ \bibnamefont
  {Cazalilla}}, \bibinfo {author} {\bibfnamefont {R.}~\bibnamefont {Citro}},
  \bibinfo {author} {\bibfnamefont {T.}~\bibnamefont {Giamarchi}}, \bibinfo
  {author} {\bibfnamefont {E.}~\bibnamefont {Orignac}}, \ and\ \bibinfo
  {author} {\bibfnamefont {M.}~\bibnamefont {Rigol}},\ }\href@noop {}
  {\bibfield  {journal} {\bibinfo  {journal} {Rev. Mod. Phys.}\ }\textbf
  {\bibinfo {volume} {83}},\ \bibinfo {pages} {1405} (\bibinfo {year}
  {2011})}\BibitemShut {NoStop}%
\bibitem [{\citenamefont {Guan}\ \emph {et~al.}(2013)\citenamefont {Guan},
  \citenamefont {Batchelor},\ and\ \citenamefont {Lee}}]{Guan-RMP13}%
  \BibitemOpen
  \bibfield  {author} {\bibinfo {author} {\bibfnamefont {X.-W.}\ \bibnamefont
  {Guan}}, \bibinfo {author} {\bibfnamefont {M.~T.}\ \bibnamefont {Batchelor}},
  \ and\ \bibinfo {author} {\bibfnamefont {C.}~\bibnamefont {Lee}},\
  }\href@noop {} {\bibfield  {journal} {\bibinfo  {journal} {Rev. Mod. Phys.}\
  }\textbf {\bibinfo {volume} {85}},\ \bibinfo {pages} {1633} (\bibinfo {year}
  {2013})}\BibitemShut {NoStop}%
\bibitem [{\citenamefont {White}(1992)}]{White-PRL92}%
  \BibitemOpen
  \bibfield  {author} {\bibinfo {author} {\bibfnamefont {S.~R.}\ \bibnamefont
  {White}},\ }\href@noop {} {\bibfield  {journal} {\bibinfo  {journal} {Phys.
  Rev. Lett.}\ }\textbf {\bibinfo {volume} {69}},\ \bibinfo {pages} {2863}
  (\bibinfo {year} {1992})}\BibitemShut {NoStop}%
\bibitem [{\citenamefont {Legeza}\ \emph {et~al.}(2003)\citenamefont {Legeza},
  \citenamefont {Roder},\ and\ \citenamefont {Hess}}]{Legeza-PRB03}%
  \BibitemOpen
  \bibfield  {author} {\bibinfo {author} {\bibfnamefont {{\"O}.}~\bibnamefont
  {Legeza}}, \bibinfo {author} {\bibfnamefont {J.}~\bibnamefont {Roder}}, \
  and\ \bibinfo {author} {\bibfnamefont {B.~A.}\ \bibnamefont {Hess}},\
  }\href@noop {} {\bibfield  {journal} {\bibinfo  {journal} {Phys. Rev. B}\
  }\textbf {\bibinfo {volume} {67}},\ \bibinfo {pages} {125114} (\bibinfo
  {year} {2003})}\BibitemShut {NoStop}%
\bibitem [{\citenamefont {Peters}\ \emph {et~al.}(2012)\citenamefont {Peters},
  \citenamefont {Kawakami},\ and\ \citenamefont {Pruschke}}]{Peters-PRL12}%
  \BibitemOpen
  \bibfield  {author} {\bibinfo {author} {\bibfnamefont {R.}~\bibnamefont
  {Peters}}, \bibinfo {author} {\bibfnamefont {N.}~\bibnamefont {Kawakami}}, \
  and\ \bibinfo {author} {\bibfnamefont {T.}~\bibnamefont {Pruschke}},\
  }\href@noop {} {\bibfield  {journal} {\bibinfo  {journal} {Phys. Rev. Lett.}\
  }\textbf {\bibinfo {volume} {108}},\ \bibinfo {pages} {086402} (\bibinfo
  {year} {2012})}\BibitemShut {NoStop}%
\bibitem [{\citenamefont {Titvinidze}\ \emph {et~al.}(2008)\citenamefont
  {Titvinidze}, \citenamefont {Snoek},\ and\ \citenamefont
  {Hofstetter}}]{Titvinidze-PRL08}%
  \BibitemOpen
  \bibfield  {author} {\bibinfo {author} {\bibfnamefont {I.}~\bibnamefont
  {Titvinidze}}, \bibinfo {author} {\bibfnamefont {M.}~\bibnamefont {Snoek}}, \
  and\ \bibinfo {author} {\bibfnamefont {W.}~\bibnamefont {Hofstetter}},\
  }\href@noop {} {\bibfield  {journal} {\bibinfo  {journal} {Phys. Rev. Lett.}\
  }\textbf {\bibinfo {volume} {100}},\ \bibinfo {pages} {100401} (\bibinfo
  {year} {2008})}\BibitemShut {NoStop}%
\bibitem [{\citenamefont {Bu\ifmmode~\check{c}\else \v{c}\fi{}a}\ \emph
  {et~al.}(2019)\citenamefont {Bu\ifmmode~\check{c}\else \v{c}\fi{}a},
  \citenamefont {Tindall},\ and\ \citenamefont {Jaksch}}]{Buca-NP19}%
  \BibitemOpen
  \bibfield  {author} {\bibinfo {author} {\bibfnamefont {B.}~\bibnamefont
  {Bu\ifmmode~\check{c}\else \v{c}\fi{}a}}, \bibinfo {author} {\bibfnamefont
  {J.}~\bibnamefont {Tindall}}, \ and\ \bibinfo {author} {\bibfnamefont
  {D.}~\bibnamefont {Jaksch}},\ }\href@noop {} {\bibfield  {journal} {\bibinfo
  {journal} {Nat. Commun.}\ }\textbf {\bibinfo {volume} {10}},\ \bibinfo
  {pages} {1730} (\bibinfo {year} {2019})}\BibitemShut {NoStop}%
\bibitem [{\citenamefont {Tindall}\ \emph {et~al.}(2019)\citenamefont
  {Tindall}, \citenamefont {Bu\ifmmode~\check{c}\else \v{c}\fi{}a},
  \citenamefont {Coulthard},\ and\ \citenamefont {Jaksch}}]{Buca-PRL19}%
  \BibitemOpen
  \bibfield  {author} {\bibinfo {author} {\bibfnamefont {J.}~\bibnamefont
  {Tindall}}, \bibinfo {author} {\bibfnamefont {B.}~\bibnamefont
  {Bu\ifmmode~\check{c}\else \v{c}\fi{}a}}, \bibinfo {author} {\bibfnamefont
  {J.~R.}\ \bibnamefont {Coulthard}}, \ and\ \bibinfo {author} {\bibfnamefont
  {D.}~\bibnamefont {Jaksch}},\ }\href@noop {} {\bibfield  {journal} {\bibinfo
  {journal} {Phys. Rev. Lett.}\ }\textbf {\bibinfo {volume} {123}},\ \bibinfo
  {pages} {030603} (\bibinfo {year} {2019})}\BibitemShut {NoStop}%
\bibitem [{\citenamefont {K{\"u}hner}\ \emph {et~al.}(2000)\citenamefont
  {K{\"u}hner}, \citenamefont {White},\ and\ \citenamefont
  {Monien}}]{Kuhner-PRB00}%
  \BibitemOpen
  \bibfield  {author} {\bibinfo {author} {\bibfnamefont {T.~D.}\ \bibnamefont
  {K{\"u}hner}}, \bibinfo {author} {\bibfnamefont {S.~R.}\ \bibnamefont
  {White}}, \ and\ \bibinfo {author} {\bibfnamefont {H.}~\bibnamefont
  {Monien}},\ }\href@noop {} {\bibfield  {journal} {\bibinfo  {journal} {Phys.
  Rev. B}\ }\textbf {\bibinfo {volume} {61}},\ \bibinfo {pages} {12474}
  (\bibinfo {year} {2000})}\BibitemShut {NoStop}%
\end{thebibliography}%

\end{document}